**Silk-Nano-Fibroin Aerogels: A Bio-Derived, Amine-Rich Platform for Rapid and Reversible $CO_2$ Capture**

Md Sariful Sheikh,[1] Lijie Guo,[2] Qiyuan Chen,[3] Bu Wang[1,*]

[1]Department of Civil and Environmental Engineering, University of Wisconsin–Madison, Madison, WI, USA
[2]Beijing General Research Institute of Mining & Metallurgy, Beijing, China
[3]Department of Materials Science and Engineering, University of Wisconsin–Madison, Madison, WI, USA

*Corresponding author. Email: bu.wang@wisc.edu

# Abstract

Despite growing interest in bio-based materials, rapid, low-temperature $CO_2$ capture using amine-rich natural sorbents has received limited attention. In recent years, various porous solid sorbents have drawn significant research interest as promising carbon capture materials. However, high synthesis cost, limited $CO_2$ adsorption capacity, sluggish adsorption-desorption kinetics, high sorbent regeneration temperature, and poor operational stability remain major challenges for their practical implementation. Here, we present silk-nano-fibroin aerogels derived from natural mulberry silk as a sustainable, amine-rich, and support-free sorbent platform for energy-efficient $CO_2$ capture. The aerogels exhibit a $CO_2$ adsorption capacity competitive with state-of-the-art amino acid and amino acid ionic liquid-based solid sorbents. Thermogravimetric analysis confirms high thermal stability up to ~250 °C—substantially higher than that of conventional amine sorbents—while complete sorbent regeneration occurs at only 60 °C. Furthermore, the silk-nano-fibroin aerogels demonstrate rapid adsorption-desorption kinetics, excellent multicycle stability, and full retention of $CO_2$ adsorption capacity under humid conditions. Spectroscopic analyses (XPS, FTIR, Raman, and solid-state $^{13}$C NMR) confirm reversible $CO_2$ chemisorption through intrinsic amine sites within the silk-fibroin backbone. Overall, this work establishes silk-nano-fibroin aerogels as a sustainable and low-cost route toward energy-efficient $CO_2$ capture.

# 1. INTRODUCTION

Carbon capture, storage, and utilization are essential strategies for managing anthropogenic $CO_2$ emissions and mitigating the adverse effects of ever-increasing atmospheric $CO_2$ levels.[1–3] Aqueous amine-based solvents have been practically used for $CO_2$ capture for several decades,[4] but their large-scale industrial deployment is limited by poor thermal stability (amine degradation at 100 – 120 °C), high regeneration temperature (≥ 120 °C), vaporization loss of amines due to their high vapor pressure, corrosion of the process equipment, and the adverse environmental impact of amine production and fugitive amines.[4–7]

Amino acid-based $CO_2$ absorbents have drawn increasing research attention due to their amine-like $CO_2$ sorption behavior, promising $CO_2$ adsorption capacity, high thermal stability, nonvolatility, and eco-friendly nature.[6,8,9] However, amino acid solution technology, like the traditional aqueous amine solvents, still has several disadvantages. The high viscosity of aqueous solvents impedes $CO_2$ sorption and desorption kinetics, restricting the practically achievable sorption capacity.[10] Additionally, the regeneration of spent aqueous solutions also demands intensive heat energy due to the high heat capacity of water, resulting in high operating costs and process-related emissions.[11,12] Alternatively, amino acid-based ionic liquids (AAILs) have demonstrated higher $CO_2$ adsorption capacity compared to pure amino acids.[13] Nonetheless, AAILs also face challenges such as inadequate adsorption capacity, high synthesis costs, and increased viscosity, restricting their viability for carbon capture.

To overcome these challenges, porous solid $CO_2$ sorbents have emerged as an energy-efficient alternative to aqueous solvents.[1,2,4,14–16] With a large surface area, a nano-porous sorbent can significantly reduce the sorbent regeneration energy cost due to the absence of water. Their highly porous structures facilitate rapid adsorption and desorption kinetics, overcoming the viscosity and contact area limitations of aqueous solvents. Additionally, solid sorbents can be regenerated through temperature or pressure swings, making them adaptable to various situations. A wide range of materials—including amine- and amino acid-functionalized porous structures such as silica, carbon-based materials, zeolites, metal-organic frameworks (MOFs), covalent organic frameworks (COFs), polymers, and composites—have been explored as potential sorbents.[1,17–20] However, challenges remain in designing optimal sorbents for large-scale implementation. High-surface-area COFs and MOFs, despite their high adsorption capacity in powder form, suffer from expensive and complex synthesis processes and significant performance loss when compressed into structured sorbents.[21–23] Carbon-based materials like graphene, carbon nanofiber, and carbon nanotubes demonstrated promising $CO_2$ capture ability, but their viability is limited due to the high cost of synthesis or poor long-term stability.[24] Zeolites exhibit high $CO_2$ adsorption capacity but require regeneration temperatures exceeding 100 °C, leading to excessive energy consumption.[25]

As an alternative solid sorbent, amino acid or AAILs grafted or impregnated porous solid supports like silica gel, MOFs, COFs, and polymers have been explored and shown high $CO_2$ adsorption capacities.[26–29] Although promising, the solid amino acid/AAIL-based sorbents need further improvements in overall synthesis cost, $CO_2$ adsorption capacity, desorption kinetics, and multicycle stability.[27,30,31] Issues such as the synthesis of high-specific surface area solid support materials, pore-blocking during amino acid grafting/impregnation, insufficient loading of amino acids, and the collapse of porous support materials during operation all impact overall performance, limiting the practical implementation of these sorbents.[26,27,32,33]

Rapid low-temperature $CO_2$ adsorption and desorption using bio-derived, amine-rich aerogels remain underexplored because existing amino acid or amine-based sorbents typically rely on solid support materials and require high-temperature regeneration. Here, we demonstrate silk-nano-fibroin aerogels that enable fast and fully reversible $CO_2$ uptake and release through intrinsic amine sites—namely, the naturally occurring amine and amide nitrogen atoms within the silk-fibroin backbone—and interconnected mesoporous channels. This design overcomes the viscosity, pore-blocking, and energy barriers typical of supported amino-acid systems, offering a sustainable pathway toward energy-efficient $CO_2$ capture using a solid sorbent.

In this work, we synthesized solid support-free silk-nanoparticles (SNPs) and silk-nano-fibroin aerogels from natural mulberry silk cocoon and evaluated their $CO_2$ capture performance for the first time. The silk-fibroin protein is a natural blend of amino acids, as represented in **Figure S1(a)**, comprising glycine (45.9 %), alanine (30.30 %), serine (12.1 %), tyrosine (5.3 %), valine (1.8 %), threonine (0.9 %), and other amino acids (3.7 %).[34] Silk has been widely utilized in diverse applications due to its low cost, biodegradability, eco-friendly nature, and intrinsic chemical stability, motivating us to explore its potential as a $CO_2$ sorbent.[35–37] In addition, silk possesses unique properties like being lightweight, thermally stable, and inherently hydrophobic, which are particularly beneficial for $CO_2$ capture applications.[38,39]

The resulting freeze-dried silk-nano-fibroin aerogels exhibit competitive $CO_2$ adsorption capacity relative to state-of-the-art amino acid and AAIL-based solid sorbents. The aerogels show excellent thermal stability (up to ~ 250 °C by thermogravimetry analysis, TGA), which is much higher than that of conventional amines. Kinetic analyses revealed rapid adsorption–desorption behavior and complete regeneration at only 60 °C, underscoring their potential for energy-efficient, low-temperature $CO_2$ capture. The aerogels also demonstrated excellent multicycle stability and retained full adsorption capacity under humid conditions. Additionally, the spent silk-based sorbents can be naturally degraded or recycled without releasing harmful chemicals. Overall, this work establishes silk-nano-fibroin aerogels as a sustainable, amine-rich, support-free sorbent platform that enables rapid and fully reversible $CO_2$ capture through intrinsic amine sites and interconnected mesoporous channels.

# 2. EXPERIMENTAL SECTION

## 2.1. Synthesis of silk-nano-fibroin aerogel from raw silk cocoon

Silk-nano-fibroin aerogel was prepared by the freeze-drying of the aqueous silk-fibroin solution and hydrogel. A schematic of the silk-nano-fibroin aerogel preparation process is demonstrated in **Figure 1**. At first, mulberry silk cocoons were cut into small pieces, boiled in a 0.05 M $Na_2CO_3$ aqueous solution for 30 min, and washed with DI water (i). The degumming of silk using 0.05 M $Na_2CO_3$ solution was repeated one more time. After removing the outer sericin layer by boiling, the silk was washed several times with cold water, and the degummed silk was air-dried. The dried silk-fibroin was dissolved in a warm aqueous 9.3 M LiBr solution at 65 °C and stirred for 6 hours (ii).[40] Finally, the silk-fibroin solution was dialyzed at 4 °C in a cellulose tube (molecular weight cut-off ~ 3.5 kDa) against water to remove LiBr (iv). To determine the weight percentage of silk-fibroin in the solution, a small amount of the measured solution was dried in an oven to evaporate the water, and the weight of the silk-fibroin was calculated to determine the weight percentage of silk-fibroin in the solution. The silk-fibroin concentration in the dialyzed solution was adjusted by adding water as required. The prepared silk-fibroin solution was stored at room temperature in an airtight plastic container. The solution generally takes 2 to 6 weeks to form the silk-fibroin hydrogel. Moreover, we observed that the silk-fibroin solution forms a hydrogel only when its weight percentage exceeds 0.2 %.

To prepare the silk-fibroin nanostructures, the silk-fibroin solution and hydrogel were frozen using the refrigerator and liquid nitrogen at -80 °C and -196 °C, respectively (vi). Finally, the frozen silk-fibroin solution and hydrogel were lyophilized at -48 °C and 0.04 mbar to obtain the silk-nano-fibroin aerogel (vii). Based on our observation, aerogel preparation from silk-fibroin solution provides a more straightforward and controllable path than aerogel preparation using silk-fibroin hydrogel, as the jellification process is time-consuming and uncontrollable.

The supporting information provides details on the SNPs synthesis, sample characterizations, $CO_2$ capture measurements, and the methods used for the adsorption-desorption kinetics study.

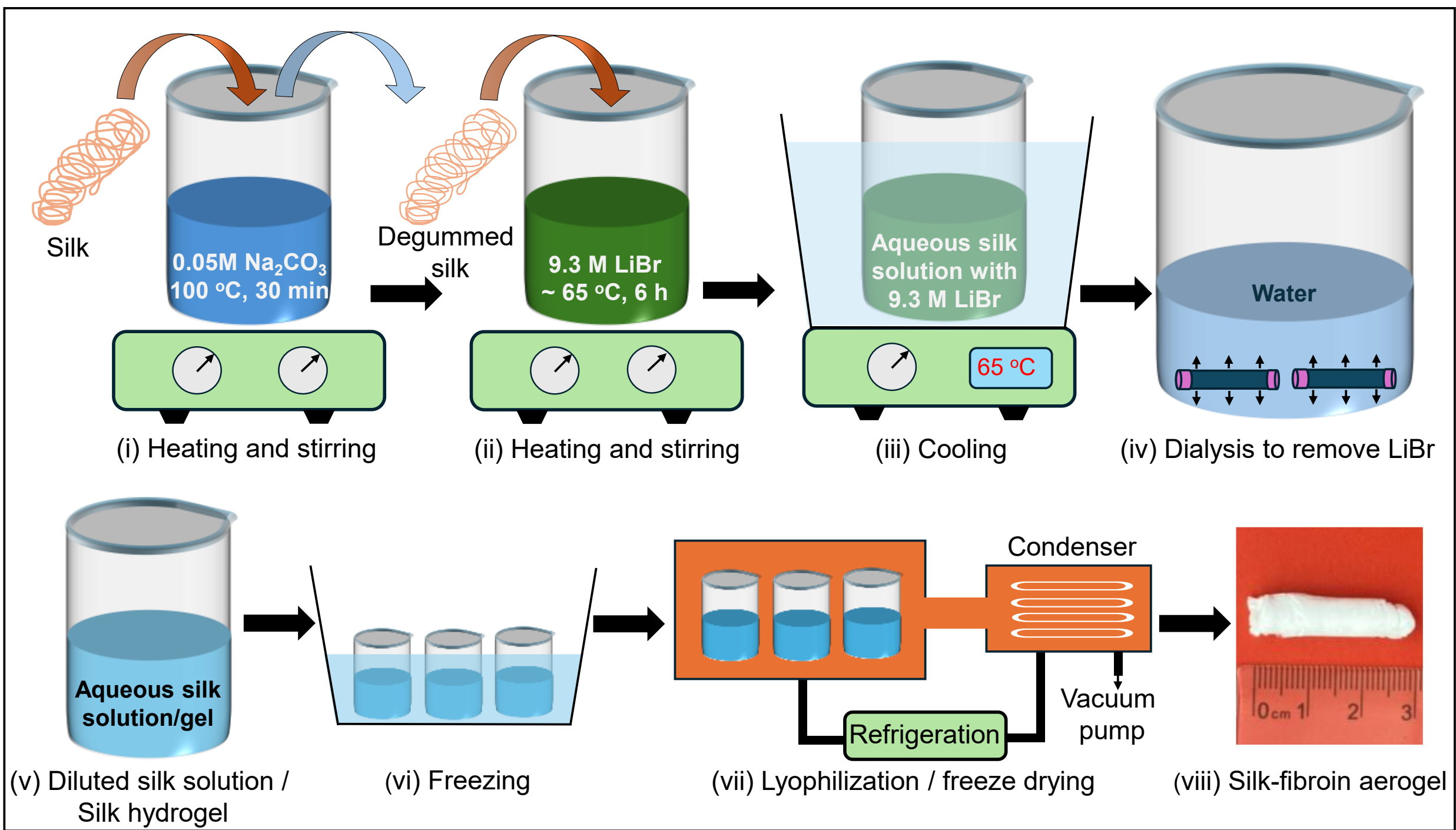

**Figure 1:** A schematic of silk-nano-fibroin aerogel preparation from aqueous silk-fibroin solution and hydrogel derived from Bombyx mori silk.

# 3. RESULTS AND DISCUSSION

To prepare the silk-fibroin-based $CO_2$ sorbent, we first attempted to prepare SNPs by partial acid hydrolysis, as represented in **Figure S1(b)** and described in **Section S1**. **Figure S2(a-f)** represents the optical images of the silk cocoon and field-effect scanning electron microscope (FESEM) images of raw silk fiber, degummed silk fiber/silk fibroin, and SNPs. The average diameter of the raw silk fiber (20 microns) is reduced to 15 microns after removing the outer sericin layer in the degummed silk. The FESEM and AFM images (**Figure S3**) confirm the submicron size of the silk particles. **Figure 2(a)** represents the high-resolution FESEM image with the roughness and porosity visible on the SNP surface. The particle size distribution was measured using Zetasizer and presented in the inset. The plot reveals two distinct particle sizes with average diameters of 140 and 460 nm. However, the yield of silk-fibroin nanoparticles prepared by this partial acid hydrolysis method was very low, and there was significant waste of silk materials.

Additionally, the performance of the nanoparticles, as demonstrated later, was limited due to the difficulty in controlling the SNP's porosity and specific surface area. To improve the $CO_2$ capture performance of silk-fibroin, we prepared silk-nano-fibroin aerogels via lyophilization of aqueous silk-fibroin solutions and hydrogels, as represented in **Figure 1**. The aqueous silk-fibroin solution and hydrogel were quickly frozen using liquid nitrogen at -196 °C (77K) and then freeze-dried (vacuum drying at -48 °C) to obtain the silk-nano-fibroin-based aerogels. The aerogel prepared using the 0.06 wt% silk-fibroin solution, identified as sol-0.06%@77K in the rest of the manuscript, exhibits a structure composed of nanosheets and nanofibers, as shown in **Figure 2(b and c)**. The aerogel prepared using the 0.25 wt% silk-fibroin hydrogel, identified as gel-0.25%@77K in the rest of the manuscript, represents nanosheet-like structures that are composed of very thin nanofibers, as represented in **Figure 2(d-f)**. The specific surface area of the SNP, sol-0.06%@77K and gel-0.25%@77K aerogels measured by the Brunauer, Emmett, and Teller (BET) method using the $N_2$ adsorption isotherm at 77 K is 104, 189, and 229 $m^2$/gm, respectively (**Figure S4**). The pore size distribution plot, as shown in **Figure S4(g, h, i)** reveals the presence of micro (< 2 nm) and meso-pores (> 2 nm) in SNP and gel-0.25%@77K, whereas sol-0.06%@77K has only meso-pores (> 2nm). Thus, the synthesis route strongly governs the resulting microstructure and porosity of the aerogels: rapid freeze-drying of dilute solution promotes open mesopores for faster gas diffusion, while denser hydrogel freezing preserves micro–mesoporous connectivity and higher specific surface area. This tunable pore hierarchy, derived directly from the silk-fibroin assembly pathway, underpins the promising $CO_2$-adsorption capacity and rapid kinetics observed in subsequent measurements.

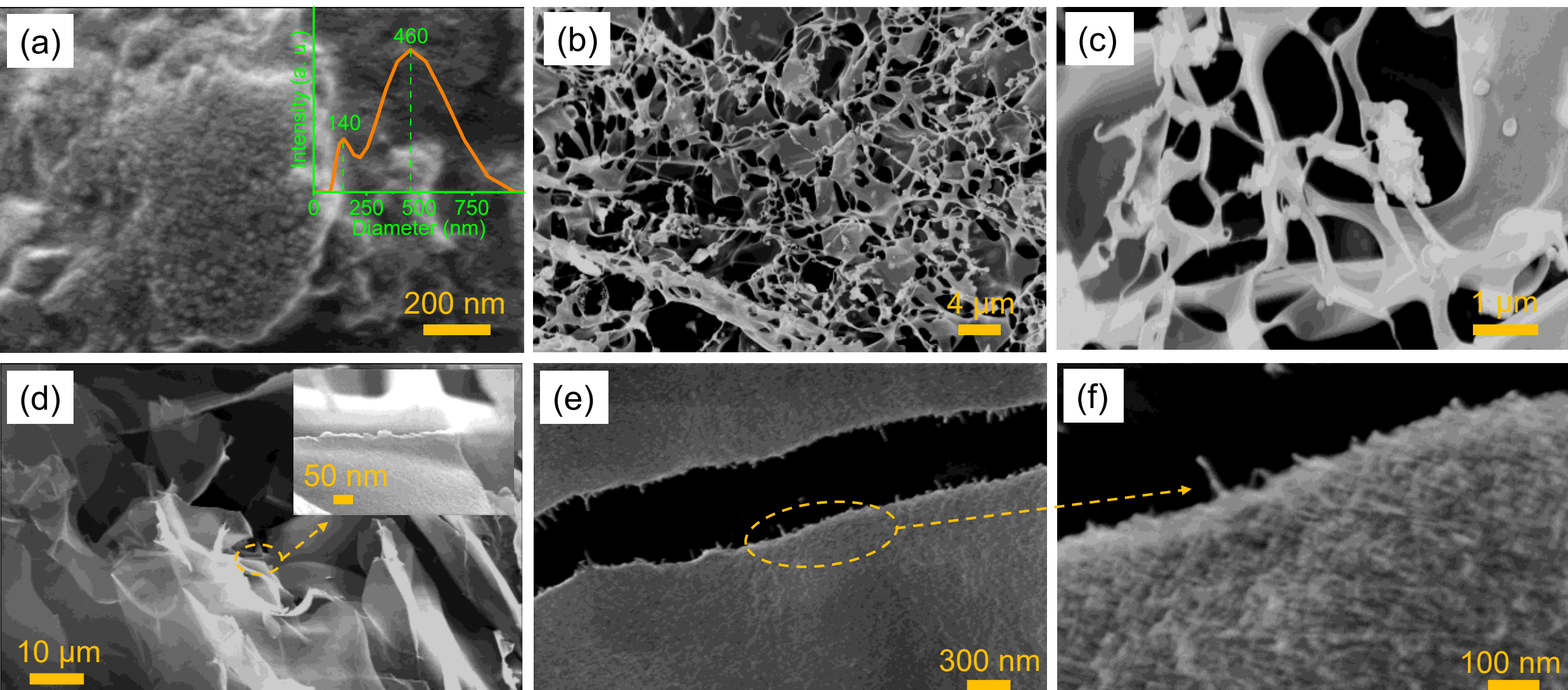

**Figure 2:** (a) FESEM image of the SNP. Inset shows their particle size distribution. (b and c) FESEM image of the silk-nano-fibroin aerogel prepared using the lyophilization of 0.06 wt% silk-fibroin solution frozen using liquid nitrogen, i.e., sol-0.06%@77K. (d-f) FESEM image of the silk-nano-fibroin aerogel prepared using lyophilization of 0.25wt% silk-fibroin hydrogel frozen using liquid nitrogen, i.e., gel-0.25%@77K.

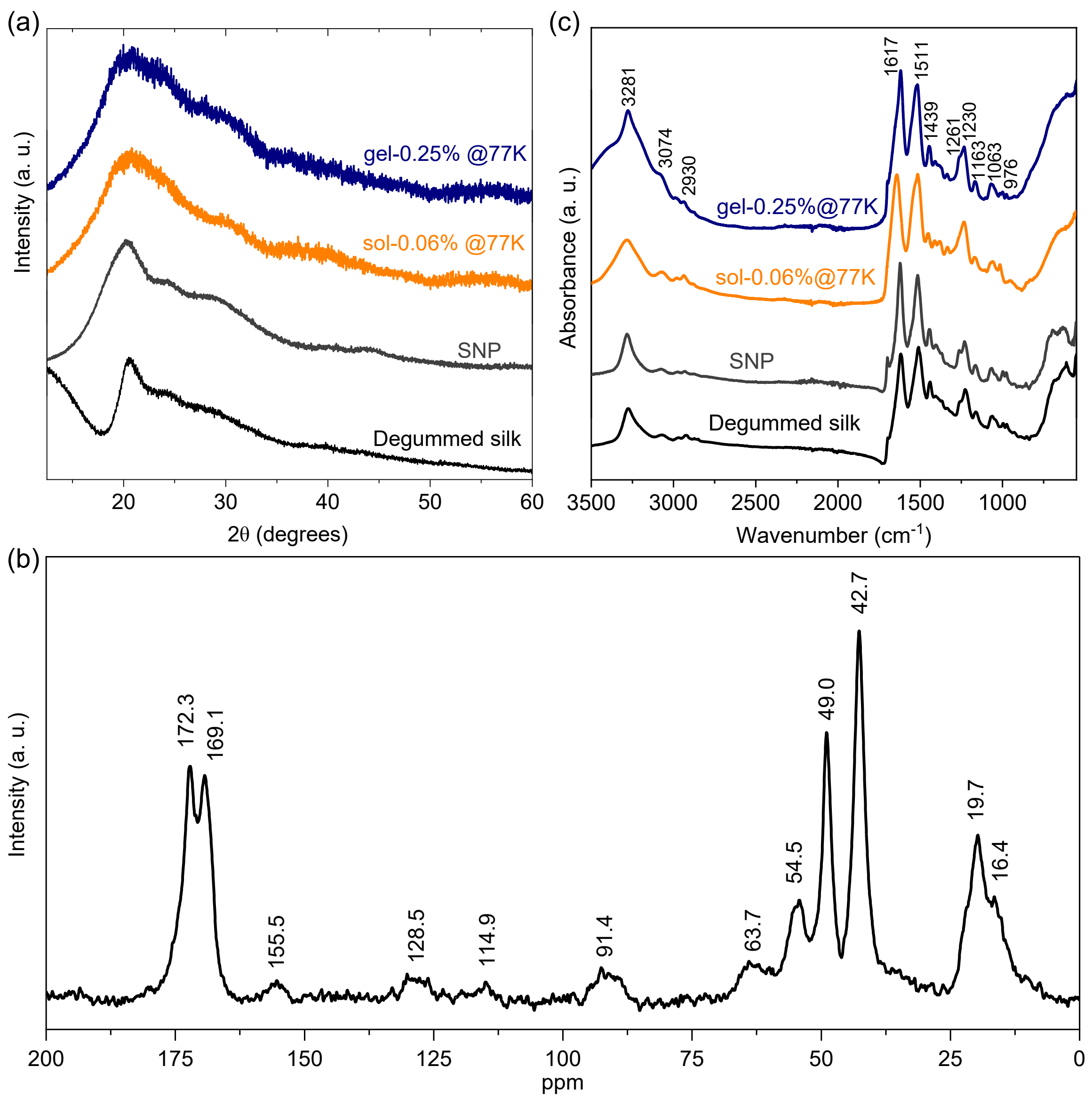

**Figure 3:** (a) Room temperature XRD patterns of the degummed silk, SNP, sol-0.06%@77K, and gel-0.25%@77K samples. (b) $^{13}C$ solid-state NMR spectra of as-prepared gel-0.25%@77K. Measurement was performed by cross-polarization (with continuous-wave decoupling of 1H). The magic angle spinning (MAS) rate was 10 kHz. (c) Fourier Transform Infrared (FTIR) spectra of the degummed silk, SNP, sol-0.06%@77K, and gel-0.25%@77K samples.

**Figure 3(a)** shows the room temperature X-ray diffraction (XRD) patterns of the degummed silk, SNP, sol-0.06%@77K, and gel-0.25%@77K aerogels. Silk fibroin has two main crystalline structures: Silk I and Silk II.[41] The peaks at around 2θ ~ 20.3° and 28.5° represent the presence of silk I structure, while the peak at around 2θ ~ 24.3° belongs to the silk II structure. However, the SNP and the aerogels do not show any significant change in the XRD peak positions with respect to the degummed silk, suggesting that the crystal structure of silk-fibroin remained unaffected after acid hydrolysis and lyophilization. Moreover, solid-state nuclear magnetic resonance (NMR) experiments were performed on as-prepared gel-0.25%@77K aerogel. The $^{13}C$ cross-polarized magic-angle spinning (CP/MAS) NMR spectrum of the silk-fibroin aerogel, as shown in **Figure 3(b)**, exhibits the significant resonances at 172.3 (Ala carbonyl carbon), 169.1 (Gly carbonyl carbon), 49.0 (Ala $C_\alpha$), 42.7 (Gly $C_\alpha$), and 19.7 (Ala $C_\beta$) ppm, which are associated with the alanine and glycine carbonyl carbon, representing the antiparallel β-sheet crystalline (silk II) structure.[42–44] A weaker shoulder peak at 16.4 ppm, besides the dominant Ala $C_\beta$, indicates the presence of a silk I or distorted β-turn domains — a coexistence commonly observed in Bombyx mori silk-fibroin. Additional peaks are observed at 155.5, 128.5, and 114.9 ppm, which are associated with aromatic carbons of Tyr, while the peaks 63.7 and 54.5 ppm arise from the Ser $C_\beta$ and $C_\alpha$ carbons, respectively. The peak around 91.4 ppm corresponds to a spinning sideband (SSB), which disappears at a spinning rate of 15K, as shown later. Overall, the $^{13}C$ NMR study confirms that silk-fibroin aerogel adopts a β-sheet crystalline (silk II) structure, with contributions from silk I or distorted β-turn domains, consistent with our XRD observations. **Figure 3(c)** shows the FTIR spectra of the degummed silk, SNP, and sol-0.06%@77K and gel-0.25%@77K, which also do not show any significant difference in the peak positions. A discussion on the observed FTIR absorption peaks is added in the supporting information file (**Section S4**). The observed absorption peaks can be attributed to amino and carboxyl groups in amino acids such as glycine and alanine. As with XRD, NMR and FTIR spectra, this suggests that the crystal structure of silk remains the same after acid hydrolysis or dissolution in salt solution. The thermal stability of the synthesized silk was examined in $N_2$, $CO_2$ and $O_2$ gas environments using thermogravimetry, as represented in **Figure S5.** The study confirms that the synthesized SNP aerogels, sol-0.06%@77K and gel-0.25%@77K, are stable up to 250 °C in inert $N_2$ and $CO_2$ atmospheres. However, the thermal degradation of silk starts at a slightly lower temperature in highly oxidizing conditions of pure $O_2$. Overall, the TGA analysis confirms the robust thermal stability of the synthesized SNP and aerogels. $CO_2$ desorption at temperatures below 100 °C is not expected to cause material degradation, as discussed later.

The effects of freezing conditions and silk-fibroin concentration on aerogel morphology were further examined using solutions at 2, 1, 0.5, and 0.25 wt%, as shown in **Figure S6**. Lower solution concentration and rapid freezing in liquid nitrogen (-196 °C) produced finer nanostructures and higher specific surface area than slower freezing at -80 °C, as shown in **Figure 4(a)**. The extremely low temperature limits molecular mobility and suppresses ice-crystal growth, yielding thin nanosheet-like frameworks instead of larger aggregated domains. Similarly, aerogels prepared from hydrogels tend to exhibit smaller structural features and higher surface areas than those from solutions of equivalent concentration, as the reduced molecular mobility in the gel phase suppresses large ice-crystal growth during freezing. However, excessive dilution reduces the synthesis yields of silk-nano-fibroin aerogels and impedes gelation, limiting practical synthesis. Accordingly, three representative systems—SNP, sol-0.06%@77K, and gel-0.25%@77K—were selected for detailed evaluation of $CO_2$-capture performance. Further optimization of freezing dynamics and concentration could enable even higher surface areas and tunable pore architectures in future work.

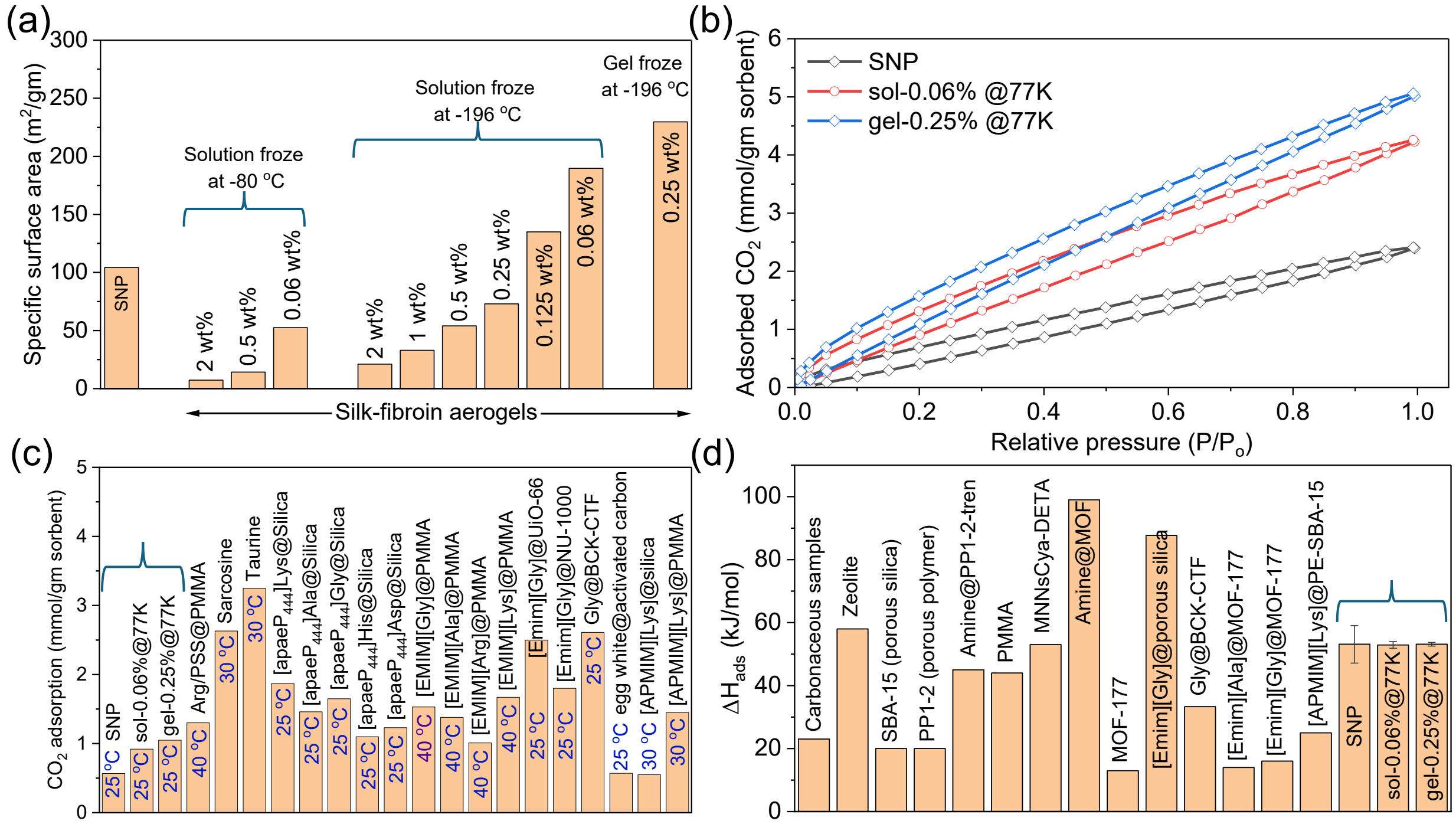

**Figure 4:** (a) Specific surface area of sorbents synthesized using partial acid hydrolysis, lyophilization of silk-fibroin solutions, and hydrogel frozen using a refrigerator (-80 °C) and liquid nitrogen (-196 °C). (b) $CO_2$ adsorption-desorption isotherms of SNP, sol-0.06%@77K and gel-0.25%@77K aerogels at 5 °C. (c) The $CO_2$ adsorption capacity (at 1 atm $CO_2$ pressure and 25 °C) of SNP, sol-0.06%@77K and gel-0.25%@77K, and its comparison with various amino acid-based solid sorbents (at 1 atm $CO_2$ pressure and near room temperature, **Table S1**).[28,30,45–51] (d) Differential heat of adsorption ($\Delta H_{ads}$) of silk-fibroin-based sorbents and their comparison with various state-of-the-art sorbents like carbonaceous samples, MOFs, zeolites at 1 mmol/gm $CO_2$ adsorption capacity (**Table S2**).[2,30,52–59]

The $CO_2$ adsorption capacity of silk-fibroin-based sorbents was measured using $CO_2$ adsorption-desorption isotherms at a temperature range from 5 to 25 °C, as demonstrated in **Figure S7(a, b, c)**. **Figure S7(d, e, f)** represents the $CO_2$ adsorption capacity as a function of temperature and pressure, determined from the $CO_2$ adsorption-desorption isotherms. The $CO_2$ adsorption capacity gradually decreases as the temperature increases and becomes poor after 25 °C. The comparison of the $CO_2$ adsorption-desorption isotherms of SNP, sol-0.06%@77K, and gel-0.25%@77K samples at 5 °C, as demonstrated in **Figure 4(b),** reveals relatively higher adsorption capacity of the aerogels compared to the SNP, which could be attributed to the higher specific surface area of the aerogels as demonstrated in **Figure 4(a)**. The measured $CO_2$ adsorption capacity of SNP, sol-0.06%@77K and gel-0.25%@77K at 1 atm $CO_2$ pressure and 25 °C was 0.57, 0.92 and 1.05 mmol/gm, respectively. Moreover, the adsorption capacity of synthesized sorbents is all competitive to the reported amino acid-based solid sorbents, including AAILs, many of which are synthesized explicitly with an increased number of $-NH_2$ groups in the molecular chain to enhance $CO_2$ adsorption, as shown in **Figure 4(c)** and **Table S1**. Most amino acid-based sorbents like taurine, sarcosine, [apaeP444][AA]@silica and [EMIM][AA]@PMMA, which demonstrated a higher adsorption capacity at nearly identical conditions, suffer from poor multicycle stability, with noticeable adsorption capacity drop after only a few cycles, as shown in **Figure S8** and **S9**. On the other hand, the multicycle stability data for the sorbents Arg/PSS@PMMA, and AAIL incorporated MOFs such as [Emim][Gly]@UiO-66 and [Emim][Gly]@NU-1000 have yet not been reported. The remaining sorbent, Gly@BCK-CTF, demonstrates promising multicycle stability, but the high specific surface area of the covalent triazine framework support material (1969 $m^2$/gm) significantly contributes to its adsorption capacity. However, silk's natural availability, low cost, facile synthesis, and biocompatibility, along with its promising multicycle stability (as discussed later), make free-standing silk-fibroin aerogels a stronger candidate compared to other solid support-based amino acid/AAIL sorbent materials. This suggests that silk-fibroin aerogel could be a potential candidate for $CO_2$ adsorption. The high absorption capacity of the aerogels may be attributed to their large specific surface area and the presence of an abundant amine group (-$NH_2$) on the surface. The performance could be further increased by enhancing their specific surface area through advanced techniques, such as $CO_2$ or $N_2$ critical point drying, instead of the conventional freeze-drying method.

The differential adsorption enthalpy, $\Delta H_{ads}$, which is an important sorbent parameter to have a quantitative understanding of the thermal energy consumption required for the sorbent regeneration, was determined from the Clausius–Clapeyron relationship using experimental isotherm (adsorption) data.[58] **Figure S10(a, b, c)** represents the $\Delta H_{ads}$ dependence of the $CO_2$ adsorption capacity of SNP, sol-0.06%@77K and gel-0.25%@77K, revealing the heterogeneity of surface energy and chemical interaction between the adsorption sites as the adsorption capacity is increased.[59] The higher $\Delta H_{ads}$ at lower capacity suggests chemisorption is a dominant mechanism followed by physisorption, as all the available amine group is saturated at higher $CO_2$ adsorption capacity. The sol-0.06%@77K and gel-0.25%@77K demonstrate a sudden drop in $\Delta H_{ads}$ value as the adsorption capacity increases above ~ 1 mmol $CO_2$/gm, suggesting the completion of the chemisorption process by the amine groups and the start of physisorption on the amine group free space on the aerogels' surface. However, we do not see a similar sudden drop in the $\Delta H_{ads}$ of SNP, possibly due to its lower adsorption capacity. From the $\Delta H_{ads}$ plots of sol-0.06%@77K and gel-0.25%@77K, we speculate that these samples can chemisorb up to around ~ 1 mmol/gm $CO_2$ by chemical interactions between the amine group of various amino acids and the $CO_2$ molecule. The average $\Delta H_{ads}$ of SNP was determined from the first 3 points, where chemisorption is hypothesized to be dominant. Similarly, for aerogels, $\Delta H_{ads}$ was estimated from the initial adsorption capacity points in **Figure S10(b, c)**, where chemisorption is dominant. The SNPs, sol-0.06%@77K and gel-0.25%@77K, show an average $\Delta H_{ads}$ value of 53.1, 52.9, and 53.1 kJ/mol, respectively, which is comparable to the reported $\Delta H_{ads}$ of the state-of-the-art solid sorbents at ~1 mmol of $CO_2$/gm capacity, as represented in **Figure 4(d)** and **Table S2**. The comparatively low $\Delta H_{ads}$ value of silk-nano-fibroin-based sorbent could be the reason behind its promising $CO_2$ adsorption. Moreover, it could be advantageous for fast $CO_2$ desorption at lower temperatures, resulting in lower energy consumption during the cyclic adsorption-desorption process by temperature swing, as demonstrated later.

We studied the multicycle stability of the aerogels' $CO_2$ adsorption capacity using 10 cycles of adsorption-desorption isotherms at 5 °C, with the results shown in **Figure 5.** To study the stability of the aerogels, the samples were tested after being exposed overnight to the laboratory air at around 22 °C and a relative humidity varying between 60% and 80 %. The $CO_2$ gas adsorption-desorption isotherms were studied on the next day after degassing the aerogel using vacuum heating at 100 °C for 30 min. Cyclic stability was studied by monitoring the adsorption capacity at 1 atm $CO_2$. The aerogel sol-0.06%@77K and gel-0.25%@77K retain their adsorption capacity after 10 cycles, showing only small fluctuations, as shown in **Figure 5(a, b)**. The promising recyclability may be attributed to its high thermal stability and low $\Delta H_{ads}$, as discussed earlier. With their high sorption capacity, silk-nano-fibroin aerogels show distinct characteristics compared to conventional amines, pure amino acids, and AAILs. In general, amines require higher desorption temperatures but have lower thermal degradation temperatures, resulting in poor cycling stability in practice, a major drawback of amine-based $CO_2$ adsorption technologies. **Figures S8 and S9** represent the $CO_2$ multicycle adsorption capacity stability of some reported amino acids and AAILs-based solid sorbents, respectively. The comparison reveals that aerogels demonstrate better cyclic stability than most of these sorbents.

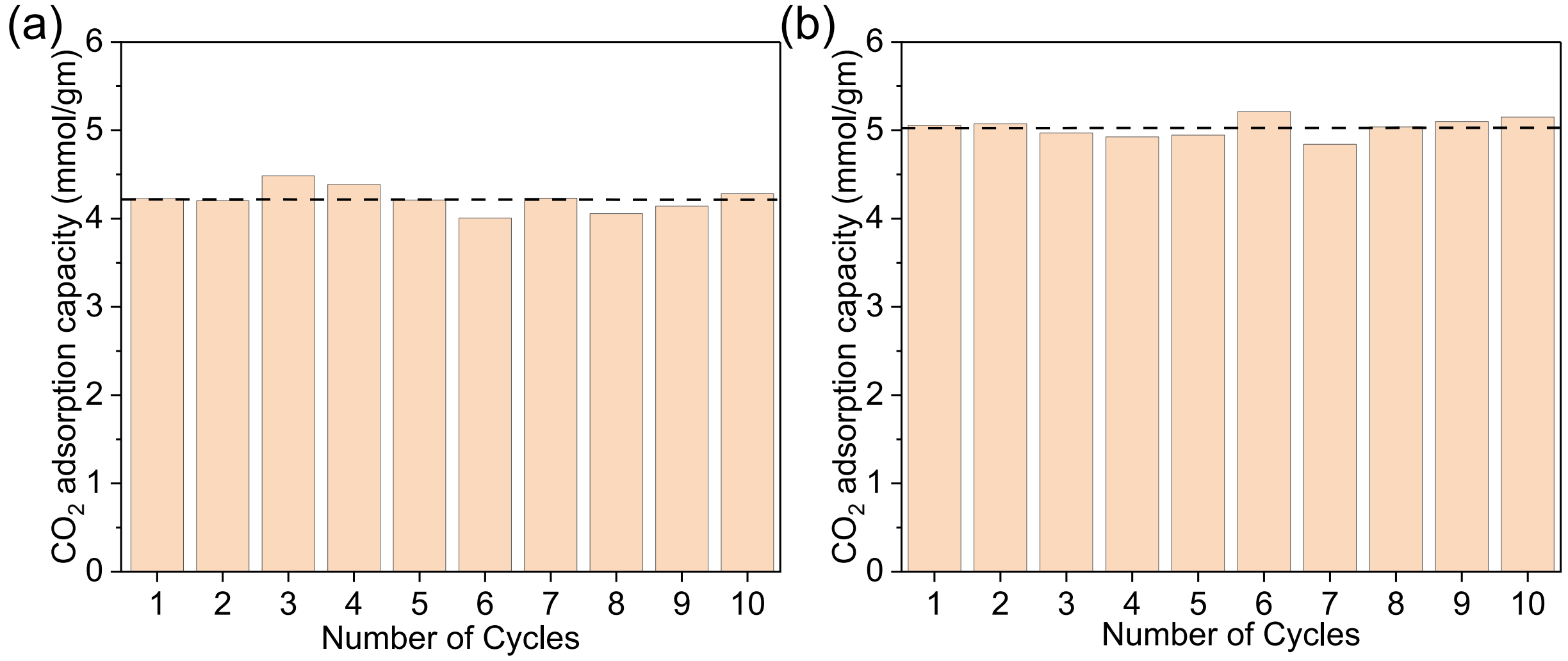

**Figure 5:** Multicycle $CO_2$ adsorption stability test of (a) sol-0.06%@77K, and (b) gel-0.25%@77K.

The effect of moisture on $CO_2$-adsorption performance was qualitatively examined to assess the aerogel's stability under humid conditions. **Figures 6(a)** and **(b)** show the multicycle $CO_2$ adsorption-desorption study in dry and humid ~13.3% $CO_2$-balanced $N_2$ gas. To study adsorption-desorption kinetics and capacity in the presence of moisture, the aerogel gel-0.25%@77K was loaded in a U-shaped quartz tube (**Figure S11**). Before each cycle, the tube was placed in a 60 °C water bath to regenerate the sample under a continuous flow of 13.3 % $CO_2$ balanced $N_2$ gas at a total flow rate of 8.3 SCCM (details in **Section S3**). After the regeneration, as the sample holder was transferred to a water bath at 5 °C, the $CO_2$ gas concentration in the outlet of the sample holder showed an immediate decrease, suggesting $CO_2$ adsorption by the aerogel. As the $CO_2$ adsorbed sample is then switched to the water bath at 60 °C, the $CO_2$ gas concentration in the sample holder outlet promptly increases, suggesting $CO_2$ release from the sorbent. The adsorption capacity in the dry and moist conditions was compared by studying their desorption at 60 °C while sending the same dry $CO_2/N_2$ gas mixture through the sample holder tube. The average of 5 desorption peak heights, when the adsorption was performed in humid conditions (83±2 % relative humidity at 5 °C), showed around a 5% increase as compared to the condition when the adsorption was performed using dry ~13.3 $CO_2$-balanced $N_2$ gas. However, the $CO_2$ adsorption kinetics of the sorbent are slightly reduced in the presence of humidity in the gas stream, as observed in the adsorption peak height difference in **Figures 6(a, b)**. **Figure 6(c)** compares the required time to complete the adsorption in dry and humid conditions, revealing slightly reduced adsorption kinetics in the presence of moisture in the adsorption gas stream. However, desorption under humid gas at 60 °C reduced the subsequent-cycle capacity, indicating that regeneration in the presence of water vapor is less efficient than under dry conditions. Overall, the silk-nano-fibroin aerogel exhibits excellent moisture tolerance, retaining its $CO_2$-adsorption capacity and fast adsorption–desorption kinetics under humid conditions, making it a promising candidate for $CO_2$ capture.

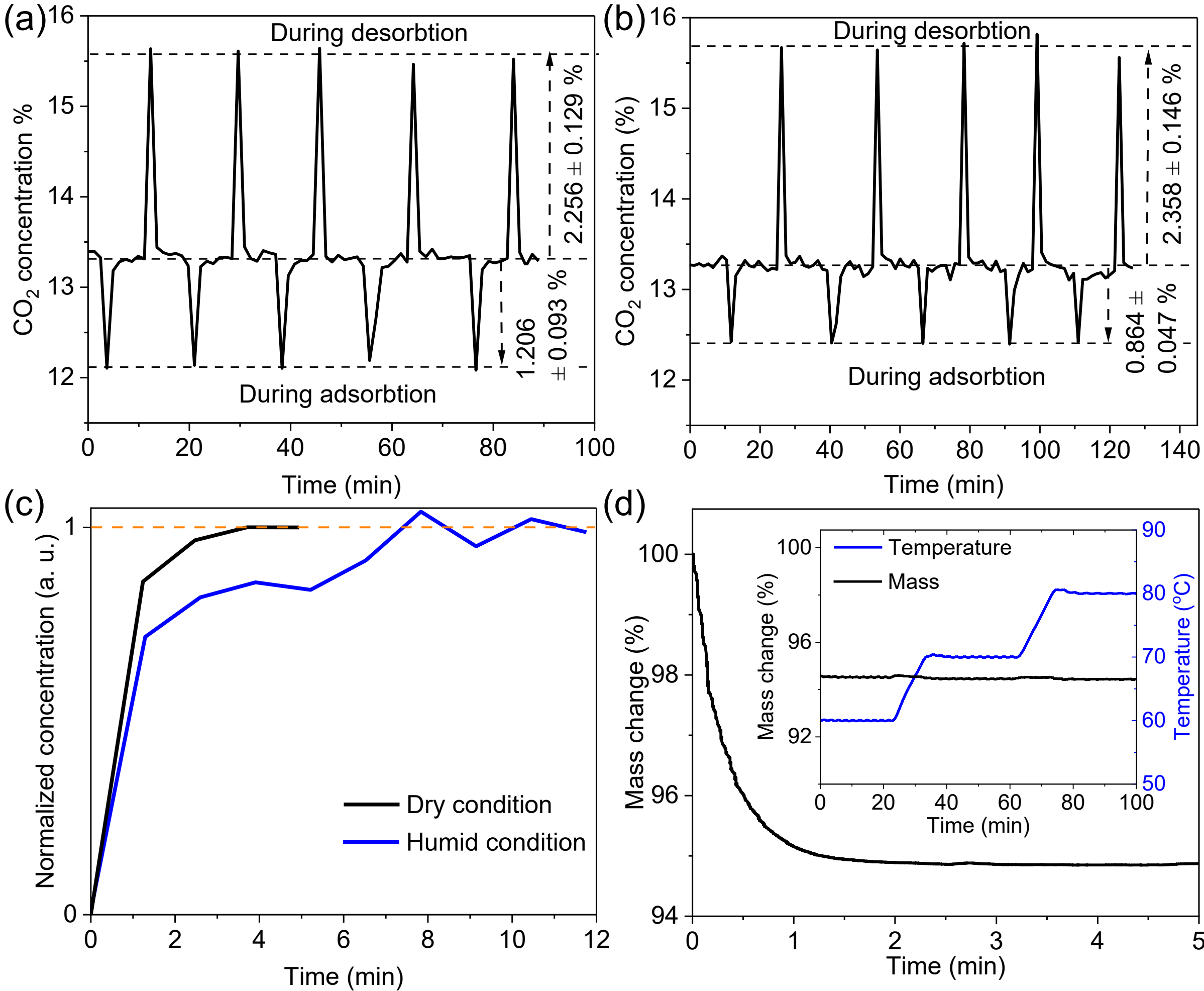

**Figure 6:** Cyclic $CO_2$ adsorption and desorption at temperatures 5 and 60 °C using ~ 13.3 % $CO_2$ balanced $N_2$ gas in (a) dry and (b) humid conditions (relative humidity 83±2% at 5 °C). (c) Normalized $CO_2$ gas concentration at the outlet of the sorbent chamber during adsorption in dry and humid conditions. (d) $CO_2$ desorption kinetics of gel-0.25%@77K at 60 °C. Inset shows the mass changes as the sample temperature increases from 60 to 80 °C stepwise.

A slow desorption rate and high temperature requirements are two major bottlenecks for implementing $CO_2$ adsorption technology. Hence, we studied qualitative $CO_2$ desorption kinetics using the aerogel gel-0.25%@77K. To study $CO_2$ desorption kinetics, $CO_2$ adsorption was first performed by exposing the aerogel to 1 atm of $CO_2$ at 23 °C for 15 minutes. Its mass change was then measured in a 1 atm $CO_2$ environment at 60 °C using TGA. **Figure 6(d)** presents the $CO_2$ desorption kinetics of gel-0.25%@77K sample at 60 °C. The sample releases all adsorbed $CO_2$ within 3 minutes, demonstrating the very fast sorbent regeneration, which can be attributed to its low heat of adsorption. We also noticed that increasing the sample temperature from 60 to 80 °C resulted in no noticeable mass change, as shown in **Figure 6(d)** inset. This suggests that sorbent regeneration was complete at 60 °C. The rapid sorbent regeneration at low temperature reveals the promising potential of self-supported silk-nano-fibroin aerogel for energy-efficient $CO_2$ capture.

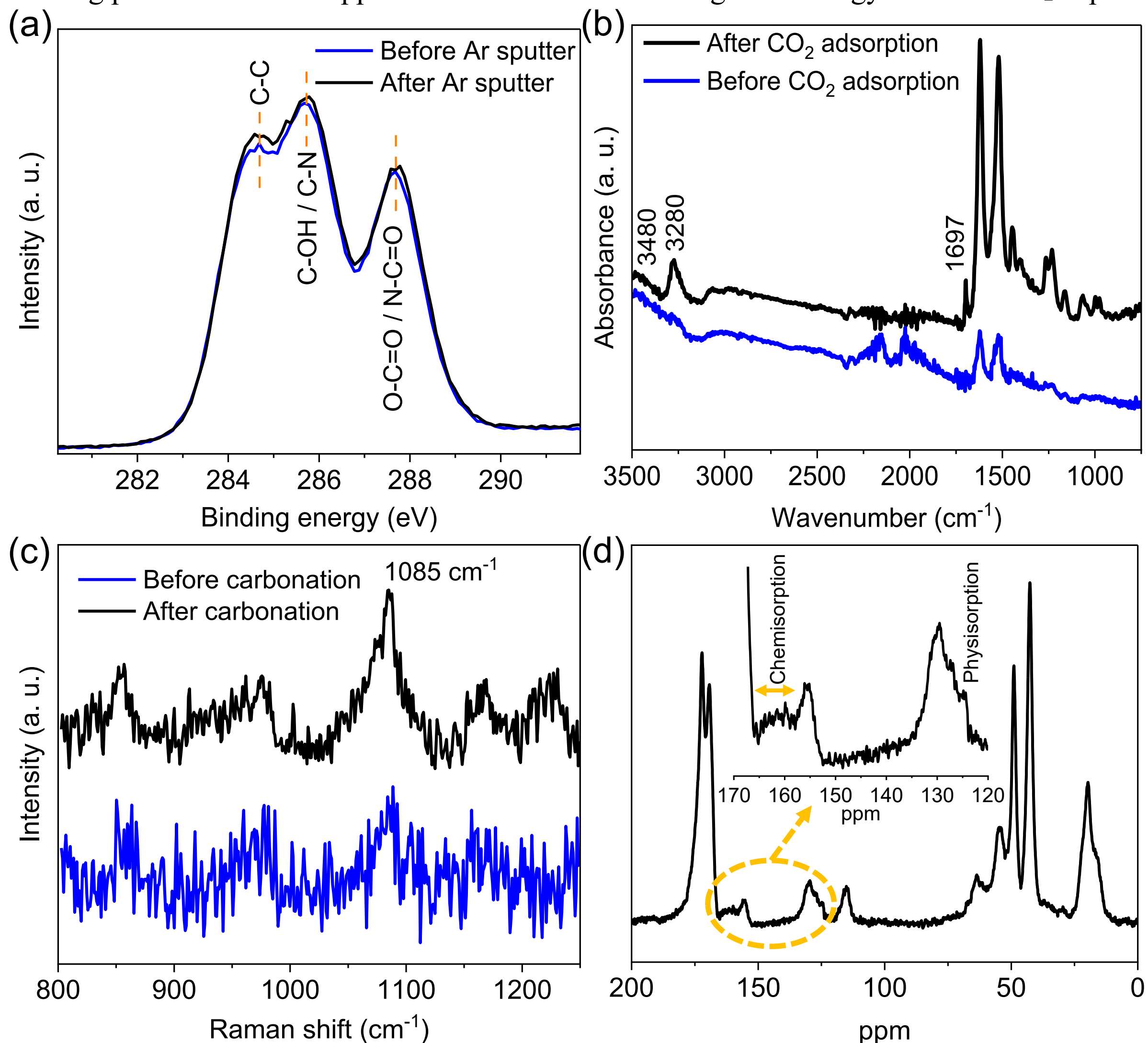

**Figure 7:** (a) Comparison of the high-resolution C-1s spectra of the $CO_2$ adsorbed sorbent before and after monoatomic $Ar^+$ ion sputtering on the sorbent surface. (b) FTIR spectra of the sorbent before and after $CO_2$ adsorption. (c) Raman spectra before and after $CO_2$ adsorption were collected in $N_2$ and $CO_2$ environments, respectively. (d) $^{13}C$ solid-state NMR spectra of silk-fibroin aerogel after $^{13}CO_2$ adsorption at atmospheric pressure and at room temperature. NMR measurements were performed using cross-polarization (with continuous-wave decoupling of 1H) with a total of 1024 scans. The magic angle spinning (MAS) rate was 15 kHz. During NMR measurement, the sample's temperature was calculated as 22.5±0.2 °C.

To elucidate the $CO_2$ adsorption mechanism on silk-nano-fibroin aerogels, X-ray photoemission spectroscopy (XPS), Raman, FTIR, and solid-state $^{13}C$ NMR spectroscopy analyses were performed. The XPS survey spectra of $CO_2$ adsorbed aerogel before and after $Ar^+$ ion sputtering (**Figure S12**) show a decrease in surface carbon (57.67 to 53.05 atomic %) and an increase in surface nitrogen (18.5 to 21.52 atomic %), indicating the removal of chemisorbed $CO_2$ and re-exposure of surface $-NH_2$ groups. The deconvolution of the core level C1s spectra, as shown in **Figure 7(a)** and **Figure S12(c, d)**, reveals a small overall increase in the C-C peak height. The fitting parameters (**Table S3**) also indicate a slight increase in the C-C bond area percentage (32.79 to 33.57 %), while the area percentage of the C-OH/C-N and O-C=O/N-C=O bonds slightly decreases, confirming the amine group-assisted $CO_2$ chemisorption on the aerogel surface.

FTIR spectra, as shown in Figure **7(b)**, further supported this interpretation. Upon $CO_2$ exposure, a new band emerges near 1697 $cm^{-1}$, characteristic of carbamate formation.[46] Additionally, the secondary amine (–NH) at 3280 $cm^{-1}$ intensified as the primary amine ($-NH_2$) group transforms to the secondary amine (–NH) upon carbamate formation, as represented in equations (1 – 3). The peak associated with the primary amine ($-NH_2$) group in $CO_2$ desorbed sample is positioned at a slightly higher wavenumber and overlapped with the broad hydroxyl group peak at 3480 $cm^{-1}$, not clearly distinguishable. These changes signify the transformation of $-NH_2$ to $-NH-COO^-$/–NHCOOH species, in agreement with previous reports.[46,60,61] Raman spectroscopy, as represented in **Figure 7(c)**, revealed an enhanced carbonate ($CO_3^{2-}$) vibration at approximately 1080 $cm^{-1}$ in the $CO_2$-adsorbed sample compared to the $CO_2$ desorbed controlled sample,[62] again confirming the chemisorption of $CO_2$ molecules on the silk-nano-fibroin aerogel surface.

To selectively probe the adsorbed carbon atoms, $^{13}CO_2$ gas was employed in solid-state $^{13}C$ cross-polarized magic angle spinning (CP/MAS) NMR. The spectrum of the silk-fibroin aerogel after $^{13}CO_2$ gas adsorption at ~ 22.5 C, as shown in **Figure 7(d)**, reveals the appearance of new resonances between 158 and 165 ppm and a sharp line at around 124.6 ppm, as compared to the as-prepared aerogel sample, as shown in **Figure 3(b)**. The 158–165 ppm region corresponds to chemisorbed $CO_2$ species (ammonium carbamate and/or carbamic acid), whereas the 124.6 ppm line arises from physisorbed or gaseous $^{13}CO_2$ confined within the pores.[63–67] Decreasing the sample temperature to around ~7 °C makes the resonance peaks more prominent with the distinct appearance of chemisorb peaks at 160.2 and 163.5 ppm and physiosorbed/gas phase peak at 124.6 ppm, as shown in **Figure S13**. The coexistence of both chemisorbed and physisorbed $CO_2$ is consistent with amine-functionalized silicas or MOFs.[63,64,66] However, in addition to the nearly overlapping feature of chemisorbed resonances, the carbon-rich background of the silk-fibroin matrix makes the overall chemisorbed $^{13}C$ peak relatively weaker than in systems where amine groups were incorporated into MOF or silica frameworks.

Collectively, XPS, FTIR, Raman, and solid-state NMR results demonstrate that $CO_2$ is chemisorbed through surface amine groups of the silk-fibroin backbone, forming carbamate and carbamic-acid species. Under dry conditions, two amine groups react with one $CO_2$ molecule to form an ammonium carbamate (Eqs. 1 – 2). Whereas, in the presence of moisture, one amine group reacts with one $CO_2$ molecule to form a bicarbonate ion (Eq. 3),[68] explaining the observed ~ 5 % increase in $CO_2$ uptake under moisture.

$$R\text{-}NH_2 + CO_2 \leftrightarrow R\text{-}NHCOOH \text{ (Carbamic acid)} \quad (1)$$

$$R\text{-}NHCOOH + R\text{-}NH_2 \leftrightarrow R\text{-}NHCOO^- \text{ (Carbamate ion)} + R\text{-}NH_3^+ \quad (2)$$

$$R\text{-}NH_2 + CO_2 + H_2O \leftrightarrow R\text{-}NH_3^+ + HCO_3^- \text{ (Bicarbonate ion)} \quad (3)$$

The intrinsic amino-acid composition of silk-nano-fibroin aerogel, which is rich in glycine, alanine, serine, and tyrosine, provides abundant surface amine sites for these reactions. Compared with state-of-the-art pure amino acids and AAIL functionalized solid sorbents, free-standing silk-nano-fibroin aerogel offers cost-effectiveness, biocompatibility, robust thermal and moisture stability, and cyclic adsorption-desorption stability. Further studies should assess its long-term operational stability under flue gas conditions, optimize the pore architecture and surface properties for enhanced high-temperature uptake, and evaluate process-level techno-economic factors to advance this natural-protein-based platform toward practical carbon-capture deployment.

# 4. CONCLUSIONS

In this study, silk-nano-fibroin aerogels derived from natural mulberry silk were developed and systematically evaluated as sustainable, amine-rich solid sorbents for $CO_2$ capture. The aerogel exhibits a high specific surface area and competitive $CO_2$ adsorption capacity, comparable to state-of-the-art amino acid-based solid sorbents. Thermogravimetric analysis confirms excellent thermal stability up to ~ 250 °C, while the material maintains full $CO_2$-uptake capacity during repeated adsorption–desorption cycles and under humid conditions. Furthermore, the aerogels demonstrate rapid adsorption–desorption kinetics and complete regeneration at only 60 °C, underscoring their potential for low-energy $CO_2$ capture. Overall, this work establishes silk-nano-fibroin aerogels as a sustainable, amine-rich, support-free sorbent platform that enables rapid and fully reversible $CO_2$ capture through intrinsic amine sites and interconnected mesoporous channels. These results highlight the promise of natural-protein-derived materials for scalable, energy-efficient carbon-capture technologies and motivate further exploration of silk-based sorbents toward practical and low-cost $CO_2$-separation processes.

# ASSOCIATED CONTENTS

The details of materials characterization methods including XRD, FTIR, solid-state NMR, AFM, FESEM, XPS, Raman spectroscopy, and gas adsorption-desorption studies are reported in the supplementary information file. The comparative table of the $CO_2$ adsorption capacity, $\Delta H_{ads}$ and XPS fitting parameters are also represented in the supporting information file.

# ACKNOWLEDGEMENTS

This material is based upon work partially supported by the National Key R&D Program of China (No. 2022YFE0135100). We also thank the support from the University of Wisconsin-Madison Office of the Vice Chancellor for Research and Graduate Education, with funding from the Wisconsin Alumni Research Foundation.

# CONFLICT OF INTEREST

The authors have no conflicts of interest to declare.

# DATA AVAILABILITY

The Figure data and raw data, as presented in the main manuscript and the supporting information, are openly available in Figshare at https://doi.org/10.6084/m9.figshare.30460787.v1.[69]

# REFERENCES

(1) Charalambous, C.; Moubarak, E.; Schilling, J.; Sanchez Fernandez, E.; Wang, J.-Y.; Herraiz, L.; Mcilwaine, F.; Peh, S. B.; Garvin, M.; Jablonka, K. M.; Moosavi, S. M.; Van Herck, J.; Ozturk, A. Y.; Pourghaderi, A.; Song, A.-Y.; Mouchaham, G.; Serre, C.; Reimer, J. A.; Bardow, A.; Smit, B.; Garcia, S. A Holistic Platform for Accelerating Sorbent-Based Carbon Capture. *Nature* **2024**, *632* (8023), 89–94. https://doi.org/10.1038/s41586-024-07683-8.

(2) Mao, H.; Tang, J.; Day, G. S.; Peng, Y.; Wang, H.; Xiao, X.; Yang, Y.; Jiang, Y.; Chen, S.; Halat, D. M.; Lund, A.; Lv, X.; Zhang, W.; Yang, C.; Lin, Z.; Zhou, H.-C.; Pines, A.; Cui, Y.; Reimer, J. A. A Scalable Solid-State Nanoporous Network with Atomic-Level Interaction Design for Carbon Dioxide Capture. *Science Advances* **2022**, *8* (31), eabo6849. https://doi.org/10.1126/sciadv.abo6849.

(3) Ragipani, R.; Sreenivasan, K.; Anex, R. P.; Zhai, H.; Wang, B. Direct Air Capture and Sequestration of CO2 by Accelerated Indirect Aqueous Mineral Carbonation under Ambient Conditions. *ACS Sustainable Chem. Eng.* **2022**, *10* (24), 7852–7861. https://doi.org/10.1021/acssuschemeng.1c07867.

(4) D'Alessandro, D. M.; Smit, B.; Long, J. R. Carbon Dioxide Capture: Prospects for New Materials. *Angewandte Chemie International Edition* **2010**, *49* (35), 6058–6082. https://doi.org/10.1002/anie.201000431.

(5) Hack, J.; Maeda, N.; Meier, D. M. Review on CO2 Capture Using Amine-Functionalized Materials. *ACS Omega* **2022**, *7* (44), 39520–39530. https://doi.org/10.1021/acsomega.2c03385.

(6) Sang Sefidi, V.; Luis, P. Advanced Amino Acid-Based Technologies for CO2 Capture: A Review. *Ind. Eng. Chem. Res.* **2019**, *58* (44), 20181–20194. https://doi.org/10.1021/acs.iecr.9b01793.

(7) Zhao, W.; Zhang, Z.; Li, Z.; Cai, N. Investigation of Thermal Stability and Continuous CO2 Capture from Flue Gases with Supported Amine Sorbent. *Ind. Eng. Chem. Res.* **2013**, *52* (5), 2084–2093. https://doi.org/10.1021/ie303254m.

(8) Hu, G.; Smith, K. H.; Wu, Y.; Mumford, K. A.; Kentish, S. E.; Stevens, G. W. Carbon Dioxide Capture by Solvent Absorption Using Amino Acids: A Review. *Chinese Journal of Chemical Engineering* **2018**, *26* (11), 2229–2237. https://doi.org/10.1016/j.cjche.2018.08.003.

(9) Ramezani, R.; Mazinani, S.; Felice, R. D. State-of-the-Art of CO2 Capture with Amino Acid Salt Solutions. *Reviews in Chemical Engineering* **2022**, *38* (3), 273–299. https://doi.org/10.1515/revce-2020-0012.

(10) Raganati, F.; Miccio, F.; Ammendola, P. Adsorption of Carbon Dioxide for Post-Combustion Capture: A Review. *Energy Fuels* **2021**, *35* (16), 12845–12868. https://doi.org/10.1021/acs.energyfuels.1c01618.

(11) Rabensteiner, M.; Kinger, G.; Koller, M.; Hochenauer, C. PCC Pilot Plant Studies with Aqueous Potassium Glycinate. *International Journal of Greenhouse Gas Control* **2015**, *42*, 562–570. https://doi.org/10.1016/j.ijggc.2015.09.012.

(12) Alivand, M. S.; Mazaheri, O.; Wu, Y.; Stevens, G. W.; Scholes, C. A.; Mumford, K. A. Development of Aqueous-Based Phase Change Amino Acid Solvents for Energy-Efficient CO2 Capture: The Role of Antisolvent. *Applied Energy* **2019**, *256*, 113911. https://doi.org/10.1016/j.apenergy.2019.113911.

(13) Saravanamurugan, S.; Kunov-Kruse, A. J.; Fehrmann, R.; Riisager, A. Amine-Functionalized Amino Acid-Based Ionic Liquids as Efficient and High-Capacity Absorbents for CO2. *ChemSusChem* **2014**, *7* (3), 897–902. https://doi.org/10.1002/cssc.201300691.

(14) Lu, C.; Zhang, X.; Chen, X. Advanced Materials and Technologies toward Carbon Neutrality. *Acc. Mater. Res.* **2022**, *3* (9), 913–921. https://doi.org/10.1021/accountsmr.2c00084.

(15) Shi, X.; Xiao, H.; Azarabadi, H.; Song, J.; Wu, X.; Chen, X.; Lackner, K. S. Sorbents for the Direct Capture of CO2 from Ambient Air. *Angewandte Chemie International Edition* **2020**, *59* (18), 6984–7006. https://doi.org/10.1002/anie.201906756.

(16) Wang, S.; Mahurin, S. M.; Dai, S.; Jiang, D. Design of Graphene/Ionic Liquid Composites for Carbon Capture. *ACS Appl. Mater. Interfaces* **2021**, *13* (15), 17511–17516. https://doi.org/10.1021/acsami.1c01242.

(17) Zentou, H.; Hoque, B.; Abdalla, M. A.; Saber, A. F.; Abdelaziz, O. Y.; Aliyu, M.; Alkhedhair, A. M.; Alabduly, A. J.; Abdelnaby, M. M. Recent Advances and Challenges in Solid Sorbents for CO2 Capture. *Carbon Capture Science & Technology* **2025**, *15*, 100386. https://doi.org/10.1016/j.ccst.2025.100386.

(18) Siegelman, R. L.; Kim, E. J.; Long, J. R. Porous Materials for Carbon Dioxide Separations. *Nat. Mater.* **2021**, *20* (8), 1060–1072. https://doi.org/10.1038/s41563-021-01054-8.

(19) Liu, H.; Zhao, W.; Lu, H.; Meng, H. Encapsulated Ionic Liquids with MOF-Driven CO2 Channels: Overcoming Kinetic Limits for Rapid Carbon Capture. *ACS Appl. Mater. Interfaces* **2025**, *17* (36), 50713–50722. https://doi.org/10.1021/acsami.5c10814.

(20) Deeg, K. S.; Damasceno Borges, D.; Ongari, D.; Rampal, N.; Talirz, L.; Yakutovich, A. V.; Huck, J. M.; Smit, B. In Silico Discovery of Covalent Organic Frameworks for Carbon Capture. *ACS Appl. Mater. Interfaces* **2020**, *12* (19), 21559–21568. https://doi.org/10.1021/acsami.0c01659.

(21) Sher, F.; Hayward, A.; Guerraf, A. E.; Wang, B.; Ziani, I.; Hrnjić, H.; Boškailo, E.; Chupin, A.; Nemţanu, M. R. Advanced Metal–Organic Frameworks for Superior Carbon Capture, High-Performance Energy Storage and Environmental Photocatalysis – a Critical Review. *J. Mater. Chem. A* **2024**, *12* (41), 27932–27973. https://doi.org/10.1039/D4TA03877K.

(22) Bui, M.; Adjiman, C. S.; Bardow, A.; Anthony, E. J.; Boston, A.; Brown, S.; Fennell, P. S.; Fuss, S.; Galindo, A.; Hackett, L. A.; Hallett, J. P.; Herzog, H. J.; Jackson, G.; Kemper, J.; Krevor, S.; Maitland, G. C.; Matuszewski, M.; Metcalfe, I. S.; Petit, C.; Puxty, G.; Reimer, J.; Reiner, D. M.; Rubin, E. S.; Scott, S. A.; Shah, N.; Smit, B.; Trusler, J. P. M.; Webley, P.; Wilcox, J.; Dowell, N. M. Carbon Capture and Storage (CCS): The Way Forward. *Energy Environ. Sci.* **2018**, *11* (5), 1062–1176. https://doi.org/10.1039/C7EE02342A.

(23) Mazaj, M.; Logar, N. Z.; Žagar, E.; Kovačič, S. A Facile Strategy towards a Highly Accessible and Hydrostable MOF-Phase within Hybrid polyHIPEs through in Situ Metal-Oxide Recrystallization. *J. Mater. Chem. A* **2017**, *5* (5), 1967–1971. https://doi.org/10.1039/C6TA10886E.

(24) Ramar, V.; Balraj, A. Critical Review on Carbon-Based Nanomaterial for Carbon Capture: Technical Challenges, Opportunities, and Future Perspectives. *Energy Fuels* **2022**, *36* (22), 13479–13505. https://doi.org/10.1021/acs.energyfuels.2c02585.

(25) Boer, D. G.; Langerak, J.; Pescarmona, P. P. Zeolites as Selective Adsorbents for CO2 Separation. *ACS Appl. Energy Mater.* **2023**, *6* (5), 2634–2656. https://doi.org/10.1021/acsaem.2c03605.
(26) Sun, L.; Gao, M.; Tang, S. Porous Amino Acid-Functionalized Poly(Ionic Liquid) Foamed with Supercritical CO2 and Its Application in CO2 Adsorption. *Chemical Engineering Journal* **2021**, *412*, 128764. https://doi.org/10.1016/j.cej.2021.128764.
(27) Mohamed Hatta, N. S.; Aroua, M. K.; Hussin, F.; Gew, L. T. A Systematic Review of Amino Acid-Based Adsorbents for CO2 Capture. *Energies* **2022**, *15* (10), 3753. https://doi.org/10.3390/en15103753.
(28) Wang, X.; Akhmedov, N. G.; Duan, Y.; Luebke, D.; Hopkinson, D.; Li, B. Amino Acid-Functionalized Ionic Liquid Solid Sorbents for Post-Combustion Carbon Capture. *ACS Appl. Mater. Interfaces* **2013**, *5* (17), 8670–8677. https://doi.org/10.1021/am402306s.
(29) Uehara, Y.; Karami, D.; Mahinpey, N. Amino Acid Ionic Liquid-Modified Mesoporous Silica Sorbents with Remaining Surfactant for CO2 Capture. *Adsorption* **2019**, *25* (4), 703–716. https://doi.org/10.1007/s10450-019-00053-1.
(30) Dong, B.; Wang, D.-Y.; Wang, W.-J.; Tian, X.-L.; Ren, G. Post Synthesis of a Glycine-Functionalized Covalent Triazine Framework with Excellent CO2 Capture Performance. *Microporous and Mesoporous Materials* **2020**, *306*, 110475. https://doi.org/10.1016/j.micromeso.2020.110475.
(31) Lyu, H.; Chen, O. I.-F.; Hanikel, N.; Hossain, M. I.; Flaig, R. W.; Pei, X.; Amin, A.; Doherty, M. D.; Impastato, R. K.; Glover, T. G.; Moore, D. R.; Yaghi, O. M. Carbon Dioxide Capture Chemistry of Amino Acid Functionalized Metal–Organic Frameworks in Humid Flue Gas. *J. Am. Chem. Soc.* **2022**, *144* (5), 2387–2396. https://doi.org/10.1021/jacs.1c13368.
(32) Jiang, B.; Wang, X.; Gray, M. L.; Duan, Y.; Luebke, D.; Li, B. Development of Amino Acid and Amino Acid-Complex Based Solid Sorbents for CO2 Capture. *Applied Energy* **2013**, *109*, 112–118. https://doi.org/10.1016/j.apenergy.2013.03.070.
(33) Huang, Z.; Karami, D.; Mahinpey, N. Study on the Efficiency of Multiple Amino Groups in Ionic Liquids on Their Sorbents Performance for Low-Temperature CO2 Capture. *Chemical Engineering Research and Design* **2021**, *167*, 198–206. https://doi.org/10.1016/j.cherd.2021.01.016.
(34) Babu, K. M. 11 - Silk Fibres – Structure, Properties and Applications. In *Handbook of Natural Fibres (Second Edition)*; Kozłowski, R. M., Mackiewicz-Talarczyk, M., Eds.; Woodhead Publishing Series in Textiles; Woodhead Publishing, 2020; pp 385–416. https://doi.org/10.1016/B978-0-12-818398-4.00013-X.
(35) Wigham, C.; Fink, T. D.; Sorci, M.; O'Reilly, P.; Park, S.; Kim, J.; Varude, V. R.; Zha, R. H. Phosphate-Driven Interfacial Self-Assembly of Silk Fibroin for Continuous Noncovalent Growth of Nanothin Defect-Free Coatings. *ACS Appl. Mater. Interfaces* **2024**, *16* (43), 58121–58134. https://doi.org/10.1021/acsami.4c07528.
(36) Oral, C. B.; Su, E.; Okay, O. Silk Fibroin-Based Multiple-Shape-Memory Organohydrogels. *ACS Appl. Mater. Interfaces* **2024**, *16* (41), 56146–56158. https://doi.org/10.1021/acsami.4c12648.
(37) Tan, X. H.; Liu, L.; Mitryashkin, A.; Wang, Y.; Goh, J. C. H. Silk Fibroin as a Bioink – A Thematic Review of Functionalization Strategies for Bioprinting Applications. *ACS Biomater. Sci. Eng.* **2022**, *8* (8), 3242–3270. https://doi.org/10.1021/acsbiomaterials.2c00313.
(38) Tao, H.; Kaplan, D. L.; Omenetto, F. G. Silk Materials – A Road to Sustainable High Technology. *Advanced Materials* **2012**, *24* (21), 2824–2837. https://doi.org/10.1002/adma.201104477.
(39) Zhou, Z.; Zhang, S.; Cao, Y.; Marelli, B.; Xia, X.; Tao, T. H. Engineering the Future of Silk Materials through Advanced Manufacturing. *Advanced Materials* **2018**, *30* (33), 1706983. https://doi.org/10.1002/adma.201706983.

(40) Rockwood, D. N.; Preda, R. C.; Yücel, T.; Wang, X.; Lovett, M. L.; Kaplan, D. L. Materials Fabrication from Bombyx Mori Silk Fibroin. *Nat Protoc* **2011**, *6* (10), 1612–1631. https://doi.org/10.1038/nprot.2011.379.

(41) Xing, L.; Wang, Y.; Cheng, J.; Chen, G.; Xing, T. Robust and Flexible Smart Silk/PEDOT Conductive Fibers as Wearable Sensor for Personal Health Management and Information Transmission. *International Journal of Biological Macromolecules* **2023**, *248*, 125870. https://doi.org/10.1016/j.ijbiomac.2023.125870.

(42) Asakura, T.; Yao, J.; Yamane, T.; Umemura, K.; Ulrich, A. S. Heterogeneous Structure of Silk Fibers from Bombyx Mori Resolved by 13C Solid-State NMR Spectroscopy. *J. Am. Chem. Soc.* **2002**, *124* (30), 8794–8795. https://doi.org/10.1021/ja020244e.

(43) Nakazawa, C. T.; Higuchi, A.; Asano, A.; Kameda, T.; Aytemiz, D.; Nakazawa, Y. Solid-State NMR Studies for the Development of Non-Woven Biomaterials Based on Silk Fibroin and Polyurethane. *Polym J* **2017**, *49* (7), 583–586. https://doi.org/10.1038/pj.2017.16.

(44) Pham, D. T.; Saelim, N.; Tiyaboonchai, W. Crosslinked Fibroin Nanoparticles Using EDC or PEI for Drug Delivery: Physicochemical Properties, Crystallinity and Structure. *J Mater Sci* **2018**, *53* (20), 14087–14103. https://doi.org/10.1007/s10853-018-2635-3.

(45) Chatterjee, S.; Rayalu, S.; Kolev, S. D.; Krupadam, R. J. Adsorption of Carbon Dioxide on Naturally Occurring Solid Amino Acids. *Journal of Environmental Chemical Engineering* **2016**, *4* (3), 3170–3176. https://doi.org/10.1016/j.jece.2016.06.007.

(46) Ren, J.; Wu, L.; Li, B.-G. Preparation and CO2 Sorption/Desorption of N-(3-Aminopropyl)Aminoethyl Tributylphosphonium Amino Acid Salt Ionic Liquids Supported into Porous Silica Particles. *Ind. Eng. Chem. Res.* **2012**, *51* (23), 7901–7909. https://doi.org/10.1021/ie2028415.

(47) Balsamo, M.; Erto, A.; Lancia, A.; Totarella, G.; Montagnaro, F.; Turco, R. Post-Combustion CO2 Capture: On the Potentiality of Amino Acid Ionic Liquid as Modifying Agent of Mesoporous Solids. *Fuel* **2018**, *218*, 155–161. https://doi.org/10.1016/j.fuel.2018.01.038.

(48) Mohamed Hatta, N. S.; Hussin, F.; Gew, L. T.; Aroua, M. K. Enhancing Surface Functionalization of Activated Carbon Using Amino Acids from Natural Source for CO2 Capture. *Separation and Purification Technology* **2023**, *313*, 123468. https://doi.org/10.1016/j.seppur.2023.123468.

(49) Huang, Z.; Mohamedali, M.; Karami, D.; Mahinpey, N. Evaluation of Supported Multi-Functionalized Amino Acid Ionic Liquid-Based Sorbents for Low Temperature CO2 Capture. *Fuel* **2022**, *310*, 122284. https://doi.org/10.1016/j.fuel.2021.122284.

(50) Xia, X.; Hu, G.; Li, W.; Li, S. Understanding Reduced CO2 Uptake of Ionic Liquid/Metal–Organic Framework (IL/MOF) Composites. *ACS Appl. Nano Mater.* **2019**, *2* (9), 6022–6029. https://doi.org/10.1021/acsanm.9b01538.

(51) Jiang, B.; Wang, X.; Gray, M. L.; Duan, Y.; Luebke, D.; Li, B. Development of Amino Acid and Amino Acid-Complex Based Solid Sorbents for CO2 Capture. *Applied Energy* **2013**, *109*, 112–118. https://doi.org/10.1016/j.apenergy.2013.03.070.

(52) Xu, C.; Bacsik, Z.; Hedin, N. Adsorption of CO2 on a Micro-/Mesoporous Polyimine Modified with Tris(2-Aminoethyl)Amine. *J. Mater. Chem. A* **2015**, *3* (31), 16229–16234. https://doi.org/10.1039/C5TA01321F.

(53) Philip, F. A.; Henni, A. Incorporation of Amino Acid-Functionalized Ionic Liquids into Highly Porous MOF-177 to Improve the Post-Combustion CO2 Capture Capacity. *Molecules* **2023**, *28* (20), 7185. https://doi.org/10.3390/molecules28207185.

(54) Fan, X.; Zhang, L.; Zhang, G.; Shu, Z.; Shi, J. Chitosan Derived Nitrogen-Doped Microporous Carbons for High Performance CO2 Capture. *Carbon* **2013**, *61*, 423–430. https://doi.org/10.1016/j.carbon.2013.05.026.

(55) Bae, T.-H.; Hudson, M. R.; Mason, J. A.; Queen, W. L.; Dutton, J. J.; Sumida, K.; Micklash, K. J.; Kaye, S. S.; Brown, C. M.; Long, J. R. Evaluation of Cation-Exchanged Zeolite Adsorbents for Post-Combustion Carbon Dioxide Capture. *Energy Environ. Sci.* **2012**, *6* (1), 128–138. https://doi.org/10.1039/C2EE23337A.
(56) Mohamedali, M.; Ibrahim, H.; Henni, A. Imidazolium Based Ionic Liquids Confined into Mesoporous Silica MCM-41 and SBA-15 for Carbon Dioxide Capture. *Microporous and Mesoporous Materials* **2020**, *294*, 109916. https://doi.org/10.1016/j.micromeso.2019.109916.
(57) Huang, Z.; Mohamedali, M.; Karami, D.; Mahinpey, N. Evaluation of Supported Multi-Functionalized Amino Acid Ionic Liquid-Based Sorbents for Low Temperature CO2 Capture. *Fuel* **2022**, *310*, 122284. https://doi.org/10.1016/j.fuel.2021.122284.
(58) Kim, E. J.; Siegelman, R. L.; Jiang, H. Z. H.; Forse, A. C.; Lee, J.-H.; Martell, J. D.; Milner, P. J.; Falkowski, J. M.; Neaton, J. B.; Reimer, J. A.; Weston, S. C.; Long, J. R. Cooperative Carbon Capture and Steam Regeneration with Tetraamine-Appended Metal–Organic Frameworks. *Science* **2020**, *369* (6502), 392–396. https://doi.org/10.1126/science.abb3976.
(59) Sheshkovas, A. Z.; Veselovskaya, J. V.; Rogov, V. A.; Kozlov, D. V. Thermochemical Study of CO2 Capture by Mesoporous Silica Gel Loaded with the Amino Acid Ionic Liquid 1-Ethyl-3-Methylimidazolium Glycinate. *Microporous and Mesoporous Materials* **2022**, *341*, 112113. https://doi.org/10.1016/j.micromeso.2022.112113.
(60) Sistla, Y. S.; Khanna, A. CO2 Absorption Studies in Amino Acid-Anion Based Ionic Liquids. *Chemical Engineering Journal* **2015**, *273*, 268–276. https://doi.org/10.1016/j.cej.2014.09.043.
(61) Wu, J.; Yang, Z.; Xie, J.; Zhu, P.; Wei, J.; Jin, R.; Yang, H. Porous Polymer Supported Amino Functionalized Ionic Liquid for Effective CO2 Capture. *Langmuir* **2023**, *39* (7), 2729–2738. https://doi.org/10.1021/acs.langmuir.2c03217.
(62) Hallenbeck, A. P.; Egbebi, A.; Resnik, K. P.; Hopkinson, D.; Anna, S. L.; Kitchin, J. R. Comparative Microfluidic Screening of Amino Acid Salt Solutions for Post-Combustion CO2 Capture. *International Journal of Greenhouse Gas Control* **2015**, *43*, 189–197. https://doi.org/10.1016/j.ijggc.2015.10.026.
(63) Fonseca, R.; Vieira, R.; Sardo, M.; Marin-Montesinos, I.; Mafra, L. Exploring Molecular Dynamics of Adsorbed CO2 Species in Amine-Modified Porous Silica by Solid-State NMR Relaxation. *J. Phys. Chem. C* **2022**, *126* (30), 12582–12591. https://doi.org/10.1021/acs.jpcc.2c02656.
(64) Forse, A. C.; Milner, P. J.; Lee, J.-H.; Redfearn, H. N.; Oktawiec, J.; Siegelman, R. L.; Martell, J. D.; Dinakar, B.; Zasada, L. B.; Gonzalez, M. I.; Neaton, J. B.; Long, J. R.; Reimer, J. A. Elucidating CO2 Chemisorption in Diamine-Appended Metal–Organic Frameworks. *J. Am. Chem. Soc.* **2018**, *140* (51), 18016–18031. https://doi.org/10.1021/jacs.8b10203.
(65) Jasil, M.; Thomas, B. Solid State NMR for Mechanistic Exploration of CO2 Adsorption on Amine-Based Silica Adsorbents. *ACS Omega* **2025**, *10* (8), 7485–7492. https://doi.org/10.1021/acsomega.4c11221.
(66) Liao, P.-Q.; Chen, X.-W.; Liu, S.-Y.; Li, X.-Y.; Xu, Y.-T.; Tang, M.; Rui, Z.; Ji, H.; Zhang, J.-P.; Chen, X.-M. Putting an Ultrahigh Concentration of Amine Groups into a Metal–Organic Framework for CO 2 Capture at Low Pressures. **2016**. https://doi.org/10.1039/C6SC00836D.
(67) Li, Y.; Duan, X.; Song, W.; Ma, L.; Jow, J. Reaction Mechanisms of Carbon Dioxide Capture by Amino Acid Salt and Desorption by Heat or Mineralization. *Chemical Engineering Journal* **2021**, *405*, 126938. https://doi.org/10.1016/j.cej.2020.126938.
(68) Uehara, Y.; Karami, D.; Mahinpey, N. Roles of Cation and Anion of Amino Acid Anion-Functionalized Ionic Liquids Immobilized into a Porous Support for CO2 Capture. *Energy Fuels* **2018**, *32* (4), 5345–5354. https://doi.org/10.1021/acs.energyfuels.8b00190.

(69) Data for Manuscript Entitled "Silk-Nano-Fibroin Aerogels: A Bio-Derived, Amine-Rich Platform for Rapid and Reversible CO2 Capture," 2025. https://doi.org/10.6084/m9.figshare.30460787.v1.

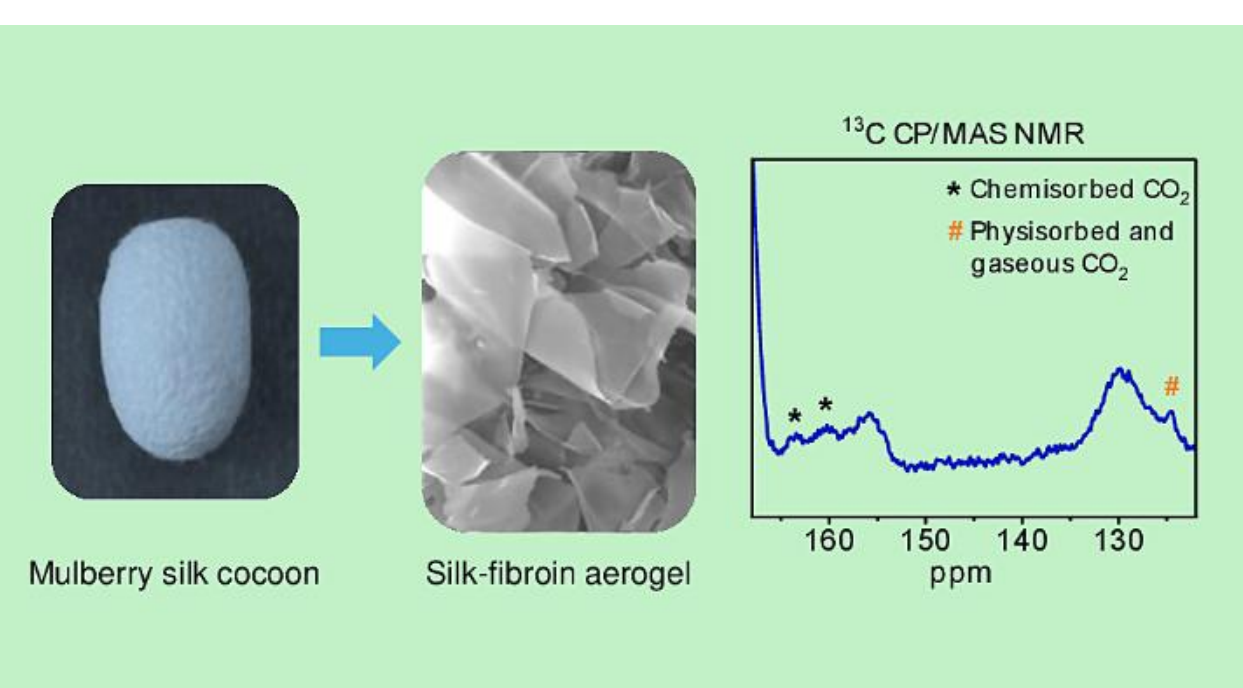

For Table of Contents Only

# Supporting Information File

## Silk-Nano-Fibroin Aerogels: A Bio-Derived, Amine-Rich Platform for Rapid and Reversible $CO_2$ Capture

Md Sariful Sheikh,[1] Lijie Guo,[2] Qiyuan Chen,[3] Bu Wang[1,*]

[1]Department of Civil and Environmental Engineering, University of Wisconsin–Madison, Madison, WI, USA
[2]Beijing General Research Institute of Mining & Metallurgy, Beijing, China
[3]Department of Materials Science and Engineering, University of Wisconsin–Madison, Madison, WI, USA

*Corresponding author. Email: bu.wang@wisc.edu

## Table of contents

### ❖ Supplementary Figures:

**Figure S9:** Multi-Cycle $CO_2$ adsorption performance of various amino acid ionic liquids (AAILs) based solid sorbents. Most of the AAILs showed gradually decreasing $CO_2$ adsorption after several cycles of adsorption and desorption. Figures are adopted with permission from the respective publishers.

**Figure S10:** Differential adsorption enthalpy ($\Delta H_{ads}$) of (a) SNP, (b) sol-0.06%@77K, and (c) gel-0.25%@77K as a function of $CO_2$ adsorption capacity.

**Figure S11:** Optical image of the gel-0.25%@77K aerogel-loaded U-shaped tube.

**Figure S12:** XPS survey scan of $CO_2$-adsrobed SNP (a) before and (b) after 10 s of monoatomic $Ar^+$ ion sputtering of energy 200 eV. Fitting of high-resolution C1s spectra (c) before and (d) after 10 s of monoatomic Ar+ ion sputtering of energy 200 eV. The symbols represent the experimental data, and the solid lines represent the fitted data.

**Figure S13:** $^{13}C$ solid-state NMR spectra of silk-fibroin aerogel after $^{13}CO_2$ adsorption. Measurement was performed by cross-polarization (with continuous-wave decoupling of $^1H$). The magic angle spinning (MAS) rate was (a, b) 10 kHz and (c, d) 15 kHz. During NMR measurement, the sample's temperature inside the rotor was calculated to be ~ 7.5 °C. The $^{13}CO_2$ gas dosing pressure was 1 bar at room temperature, though it is expected to be lower at the measured sample temperature due to pressure drop from additional adsorption and temperature drop. The number of scans was 1024 in each case.

## ❖ Supplementary discussion:

## ❖ Supplementary Tables:

**Table S1:** Comparison of the $CO_2$ adsorption capacity of silk-nano-fibroin-based sorbents with other high-performing amino acid and AAILs-based solid sorbents reported at nearly similar conditions.

**Table S2:** Comparison of differential adsorption enthalpy ($\Delta H_{ads}$) of silk-nano-fibroin-based with other state-of-the-art sorbents reported at 1 mmol $CO_2$/gm sorbent adsorption capacity.

**Table S3:** Fitting parameter of C1s core-level spectra before and after $Ar^+$-ion sputtering.

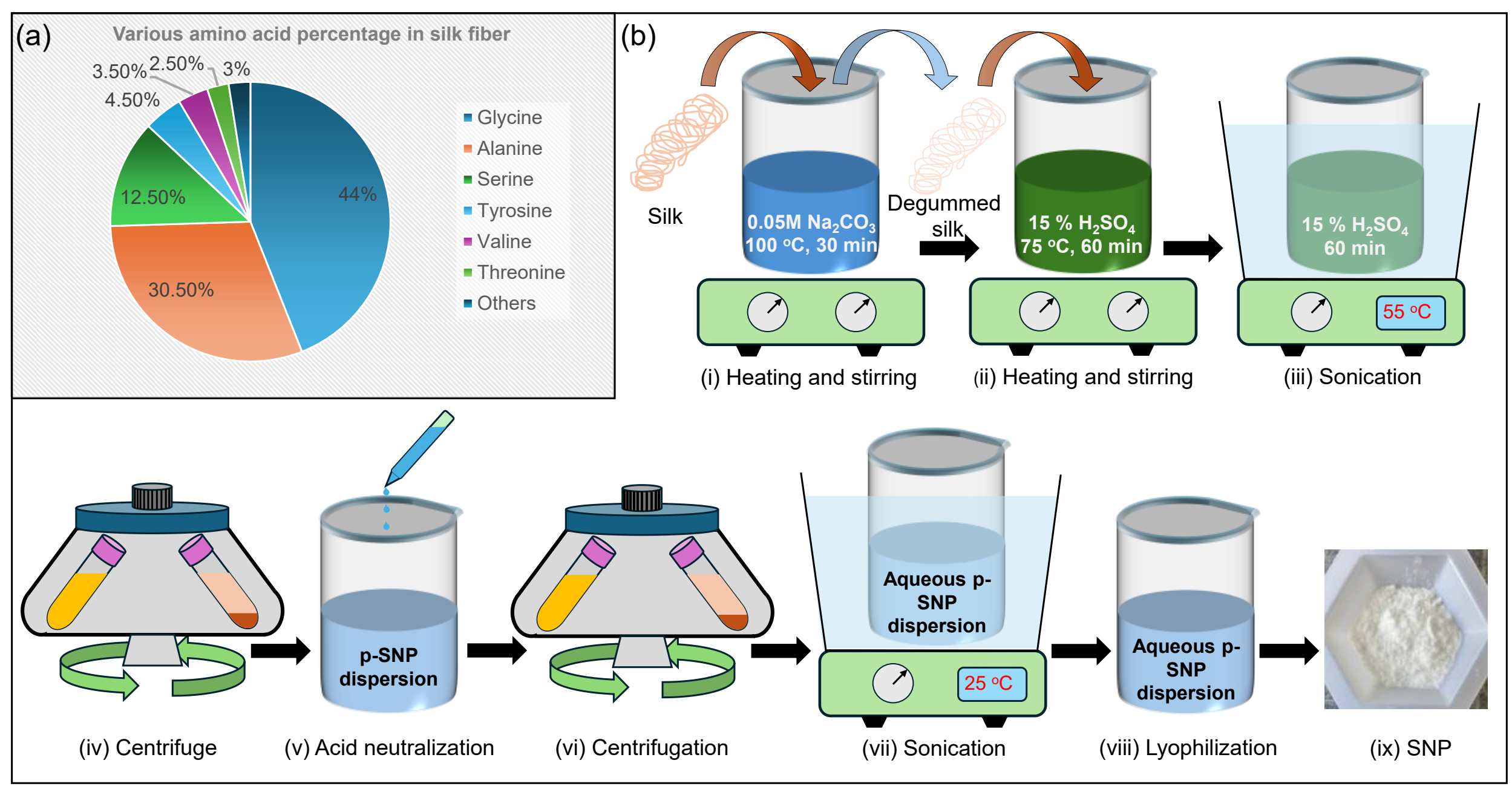

**Figure S1:** (a) Percentage of various amino acids in the silk-fibroin[1]. (b) A schematic of the porous silk-nanoparticles (SNP) synthesis from mulberry silk cocoon.

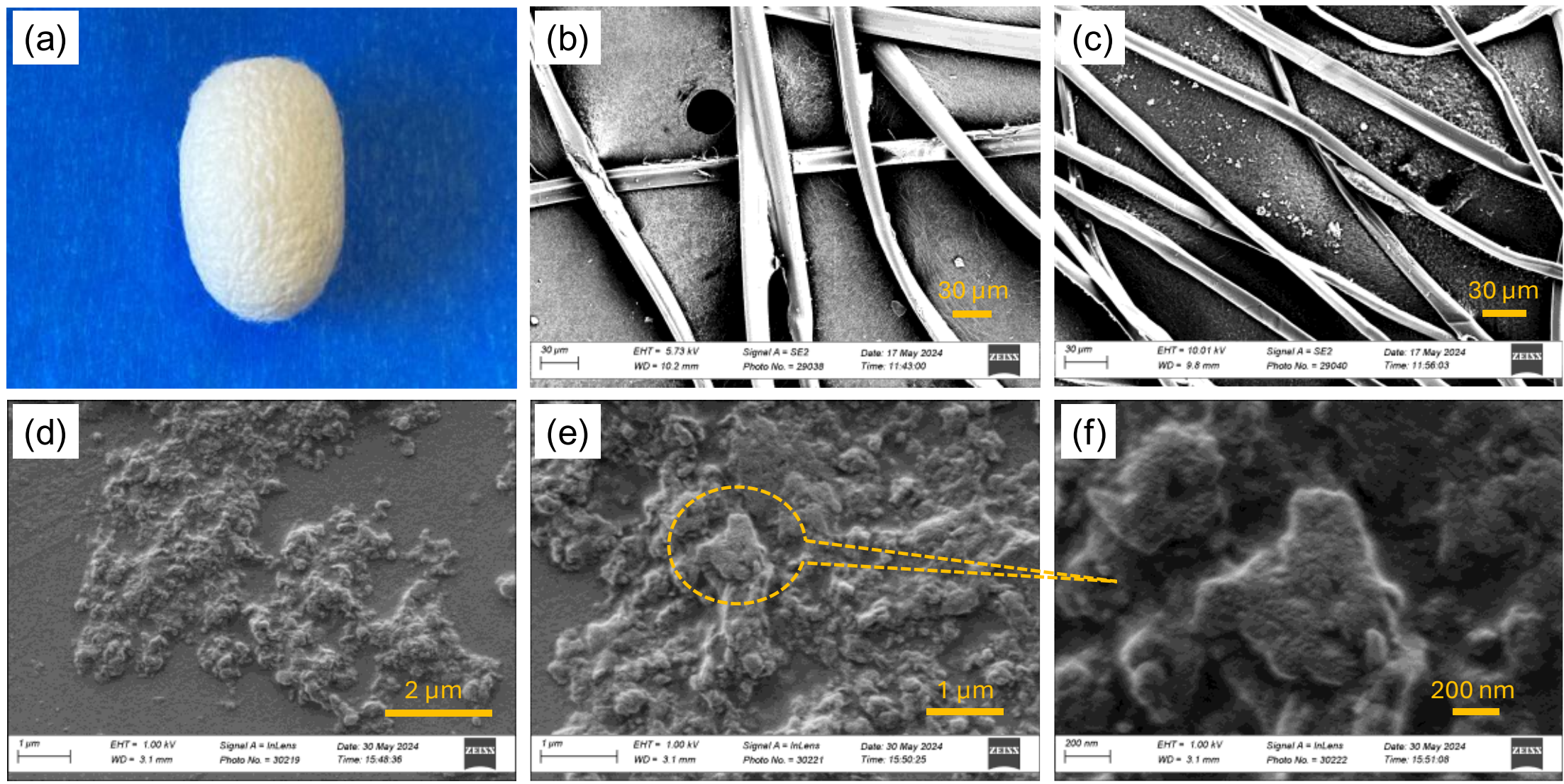

**Figure S2:** (a) Optical image of the mulberry silk cocoon. (b) Field effect scanning electron microscope (FESEM) image of the raw silk fiber. (c) FESEM image of the degummed silk fiber. (d, e, f) FESEM image of the SNP at various resolutions.

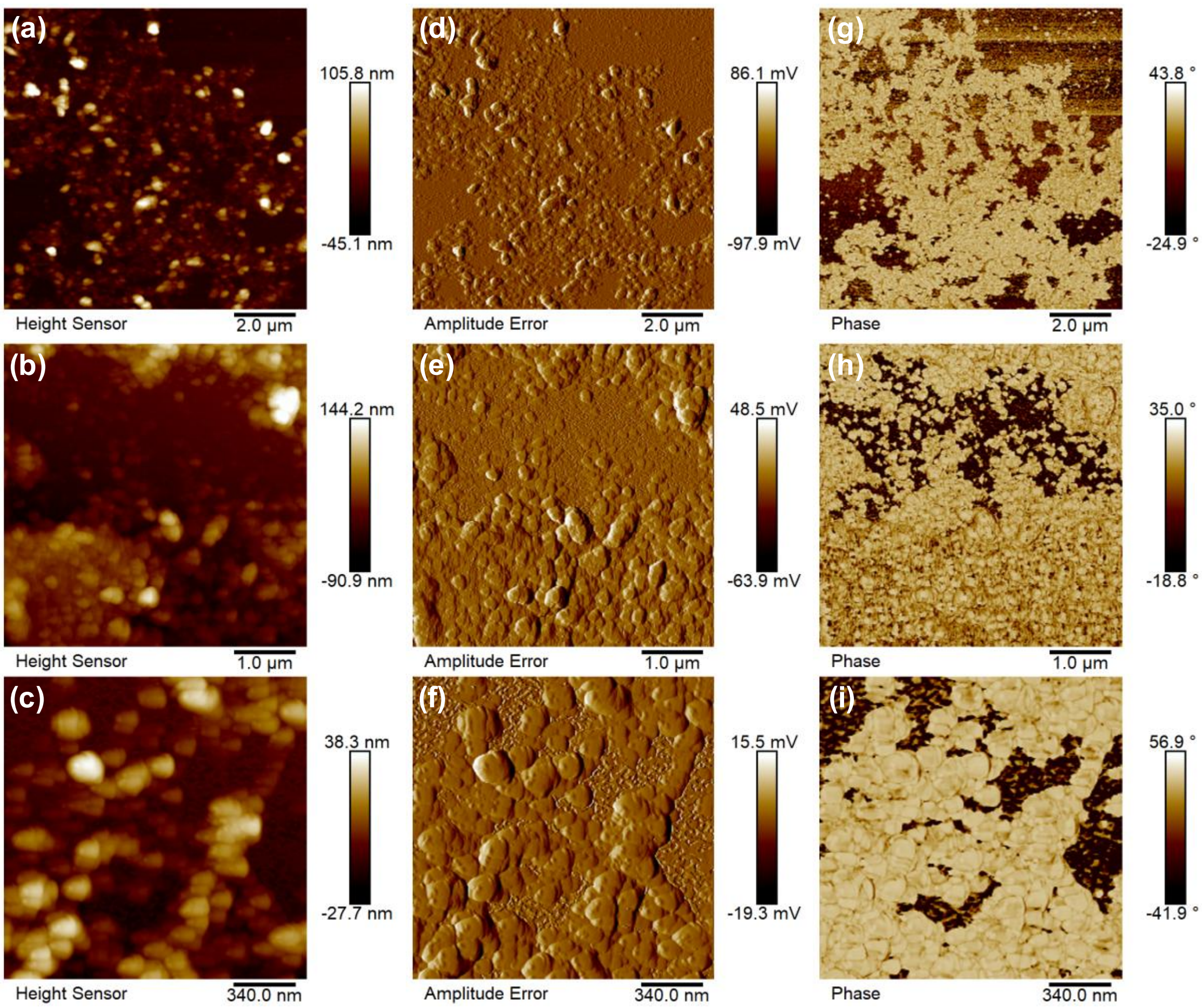

**Figure S3:** AFM (a, b, c) topography image; simultaneously acquired (d, e, f) amplitude; and (g, h, i) phase image of the SNP.

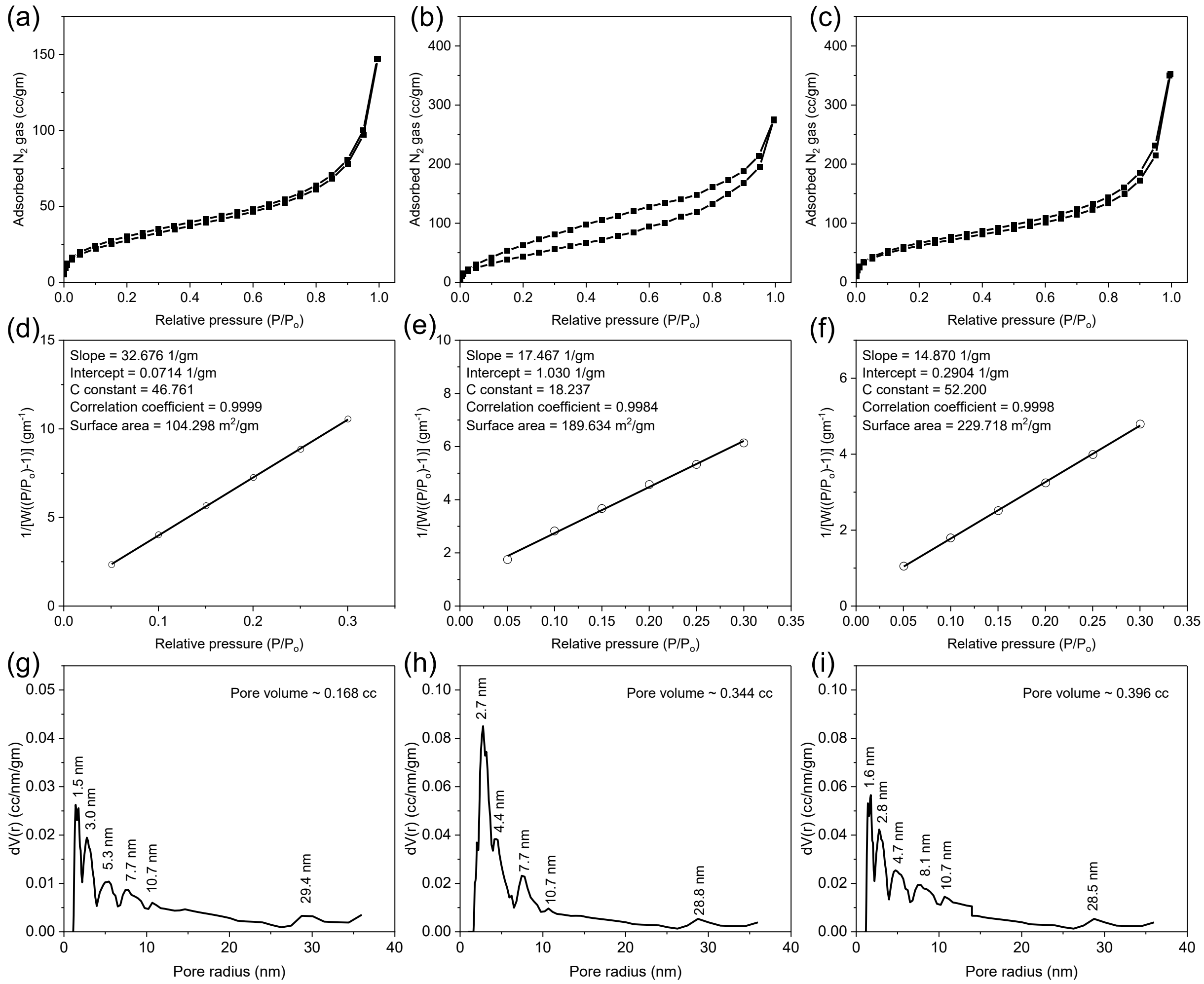

**Figure S4:** $N_2$ adsorption-desorption isotherm of (a) SNP, (b) sol-0.06%@77K, and (c) gel-0.25%@77K aerogels at 77 K. Specific surface area analysis of (d) SNP, (e) sol-0.06%@77K, and (f) gel-0.25%@77K using the Brunauer–Emmett–Teller (BET) method. Pore size distribution plot of (g) SNP, (h) sol-0.06%@77K, and (i) gel-0.25%@77K samples measured using density functional theory (DFT) pore size distribution analysis.

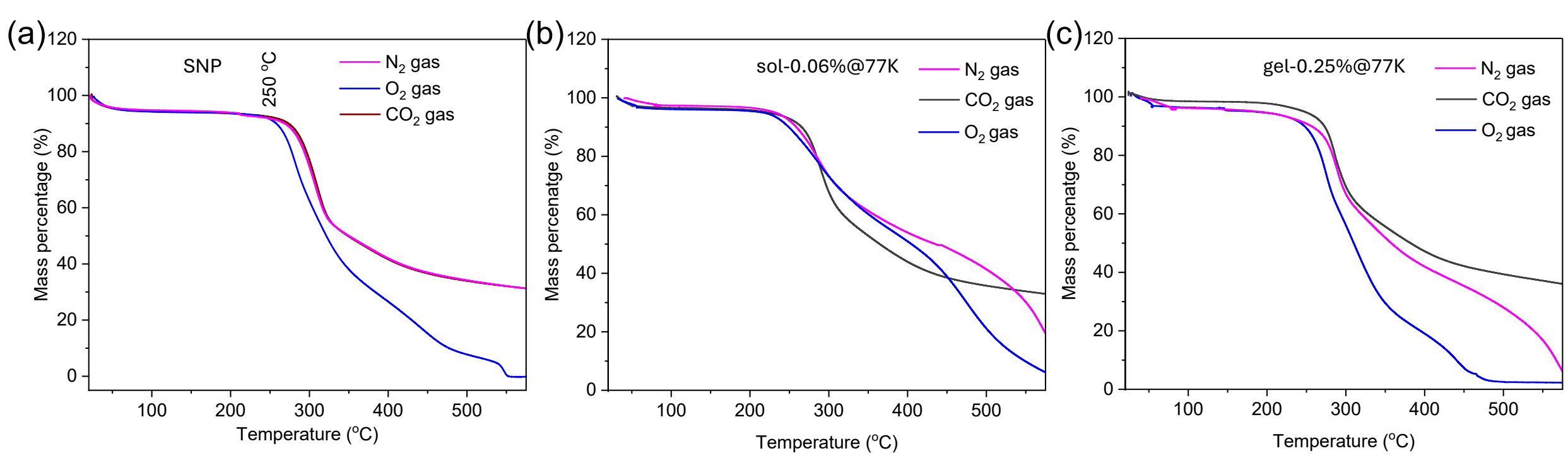

**Figure S5:** Thermal stability test of (a) SNP, (b) sol-0.06%@77K aerogel, and (c) gel-0.25%@77K aerogel samples in $O_2$, $CO_2$ and $N_2$ environments using thermogravimetry. The heating rate was 5 °C/min in all cases.

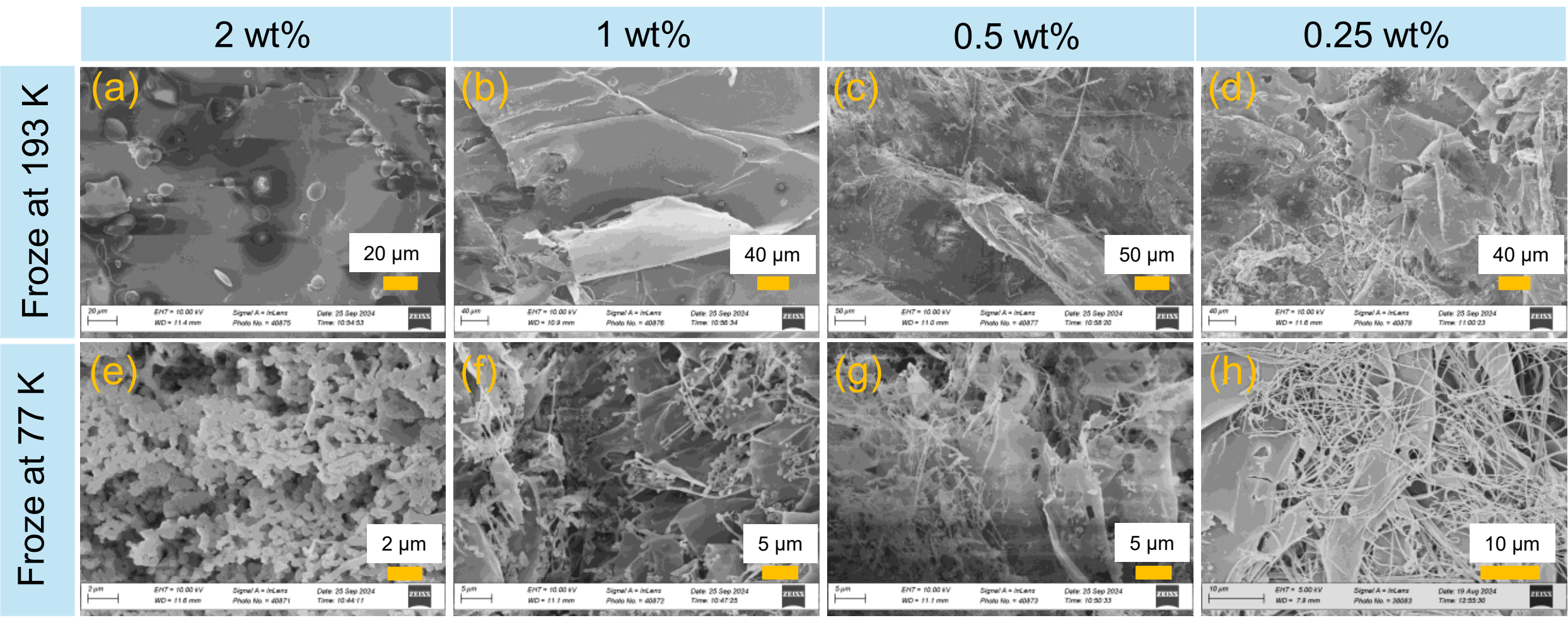

**Figure S6:** FESEM images of silk-fibroin aerogel prepared using lyophilization of 2, 1, 0.5 and 0.25 wt% aqueous silk solution froze using (a-d) refrigerator at -80 °C (193 K), and (e-h) liquid nitrogen at -196 °C (77 K).

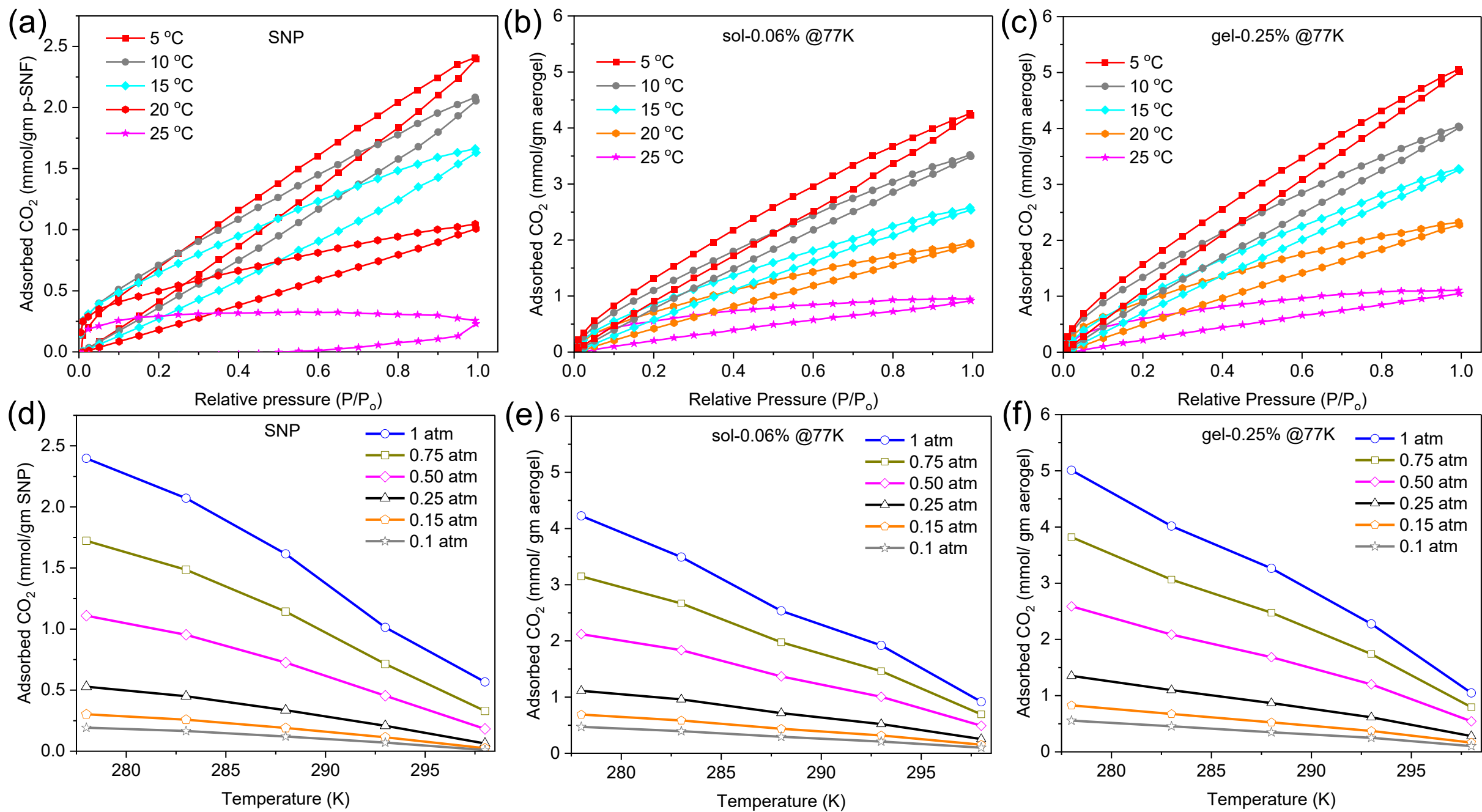

**Figure S7:** $CO_2$ adsorption-desorption isotherms of (a) SNP, (b) sol-0.06%@77K, and (c) gel-0.25%@77K at various temperatures. Temperature-dependent $CO_2$ adsorption capacity of (d) SNP, (e) sol-0.06%@77K, and (f) gel-0.25%@77K as a function of $CO_2$ pressure.

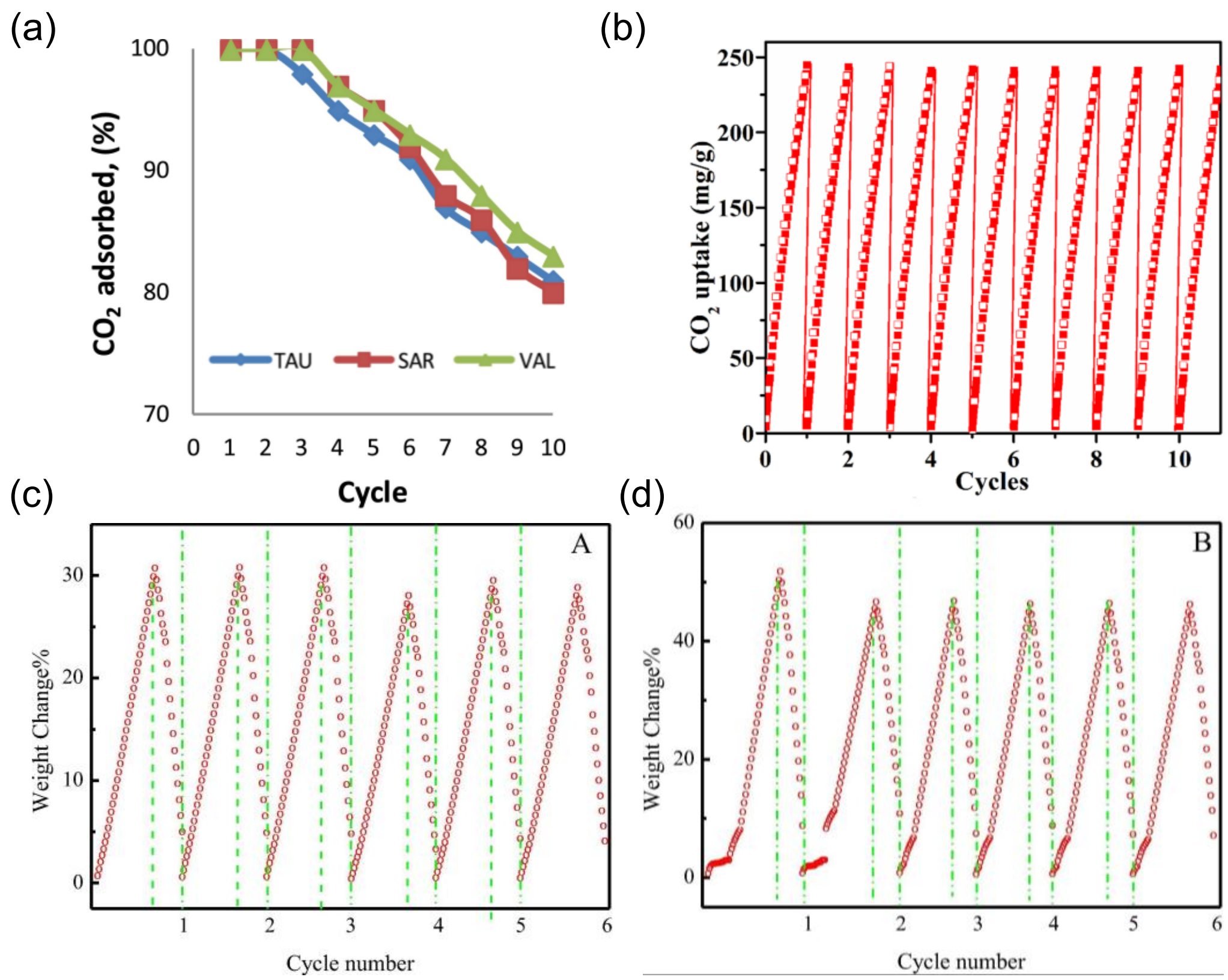

**Figure S8:** Multi-Cycle $CO_2$ adsorption performance of various amino acid-based solid sorbents. Figures are adopted with permission from the respective publishers.

(a) Around 20 % drop in adsorption capacity after 10 cycles of adsorption-desorption test using porous taurine, sarcosine and valine[2].

(b) Glycine functionalized covalent triazine framework (BCK-CTF) showed stable $CO_2$ adsorption capacity in 10 cycles[3].

(c, d) Hyper crosslinked polymer (HCP), HCP(St-DMDAAC) and glycine incorporated HCP(St-DMDAAC) adsorbents showed 2.7% and a 5% drop respectively, in the $CO_2$ adsorption capacity after 6 cycles[4].

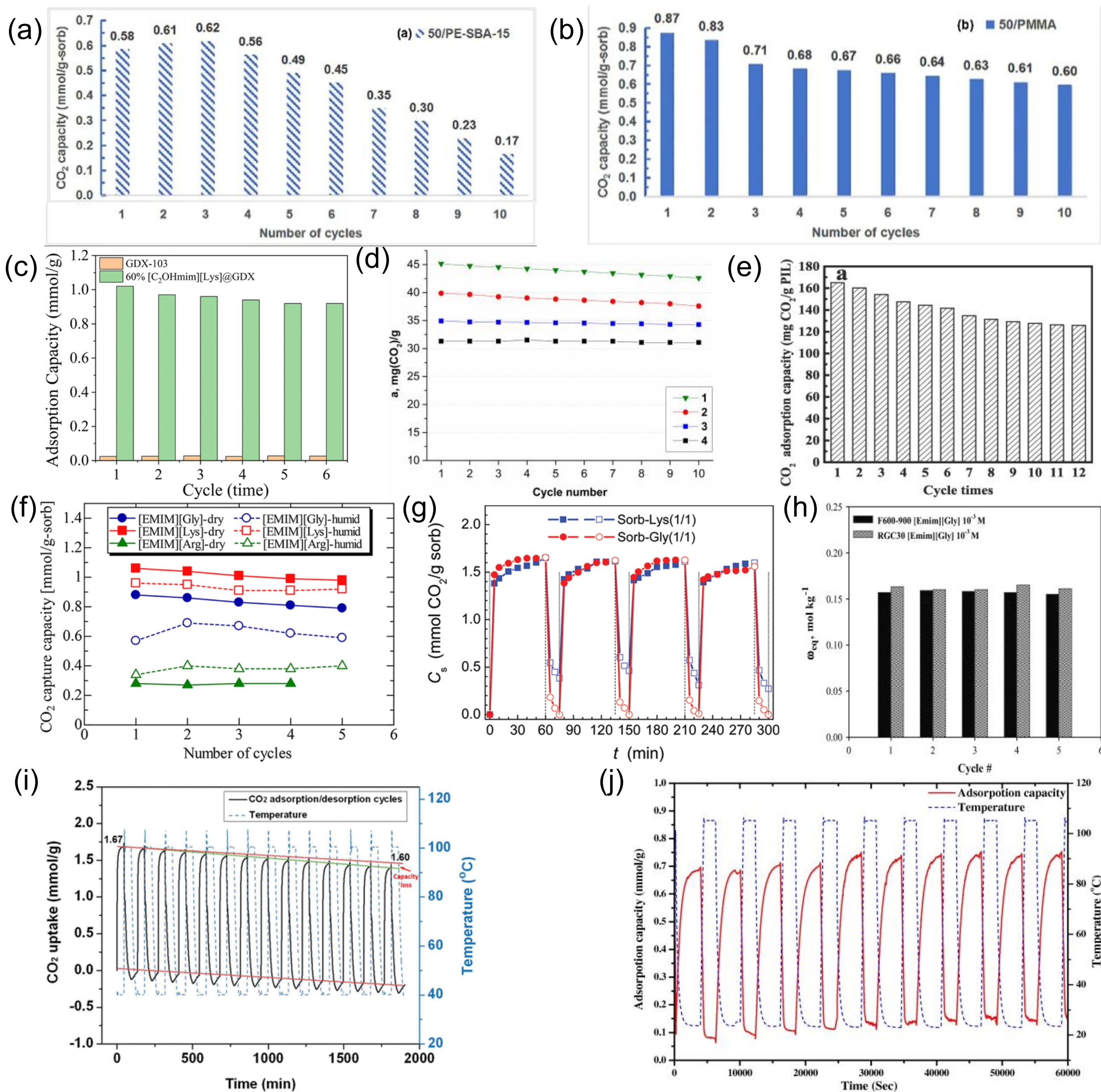

**Figure S9:** Multi-Cycle $CO_2$ adsorption performance of various amino acid ionic liquids (AAILs) based solid sorbents. Most of the AAILs showed gradually decreasing $CO_2$ adsorption after several cycles of adsorption and desorption. Figures are adopted with permission from the respective publishers.

(a, b) Aminoethyl-3-methylimidazolium Lysine, [AEMIM][Lys] functionalized mesoporous silica and poly(methyl methacrylate) (PMMA) shows gradual decrease in $CO_2$ adsorption capacity during cyclic adsorption desorption study.[5]

(c) Amino functionalized ionic liquid [$C_2$OHmim][Lys] impregnated on a chromatographic column filler poly-divinylbenzene porous spheres, GDX-103 shows gradual decrease in $CO_2$ adsorption capacity[6].

(d) 1-ethyl-3-methylimidazolium glycinate [EMIM][Gly] loaded mesoporous silica gel shows gradually decreasing $CO_2$ adsorption capacity. According to this study high high-temperature regeneration helps in better adsorption capacity, but degrades the sample faster[7].

(e) Porous poly[1-(p-vinylbenzyl)-3-methylimidazolium glycinate], [P([VBMI][Gly])] sorbent shows gradual decrease in $CO_2$ adsorption capacity due to the destruction of its pore structure during the regeneration process[8].

(f) Glycine and lysine based AAILs functionalized poly(methyl methacrylate) (PMMA) show gradually decreasing $CO_2$ adsorption capacity in both dry and humid conditions. Arginine based AAIL shows nearly stable $CO_2$ adsorption in both the conditions. However, [EMIM][Arg] has a very poor adsorption capacity [9].

(g) N-(3-aminopropyl)aminoethyl tributylphosphonium amino acid, [apaeP$_{444}$][AA] impregnated mesoporous silica sorbents show gradual decrease in $CO_2$ adsorption capacity[10].

(h) [Emim][Gly] AAIL functionalized activated carbon have nearly constant $CO_2$ adsorption capacity during 4 cycle tests. However, the sorbent has very poor adsorption capacity[11].

(i) 4% loss in $CO_2$ adsorption capacity after 14 cycles for [EMIM][Lys] impregnated PMMA (48.7 wt %) sorbent. [EMIM][Lys] has a thermal degradation temperature of around 200 °C[12].

(j) Nearly stable $CO_2$ adsorption capacity observed after 10 cycles of adsorption-desorption test using 1-methyl-3-ethyl-imidazolium lysinate (OMS-IL-Lys) grafted mesoporous silica. However, the sorbent requires a high regeneration temperature of 105 °C due to its high value of differential heat of adsorption ($\Delta H_{ads}$ ~ 85.7 kJ/mol) [13].

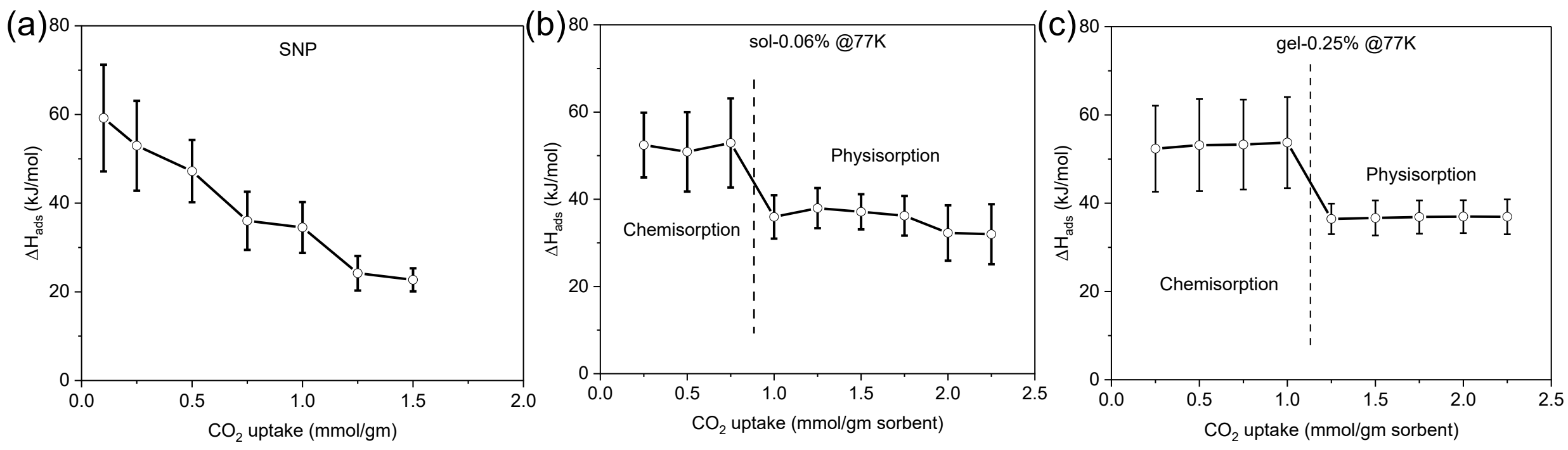

**Figure S10:** Differential adsorption enthalpy ($\Delta H_{ads}$) of (a) SNP, (b) sol-0.06%@77K, and (c) gel-0.25%@77K as a function of $CO_2$ adsorption capacity.

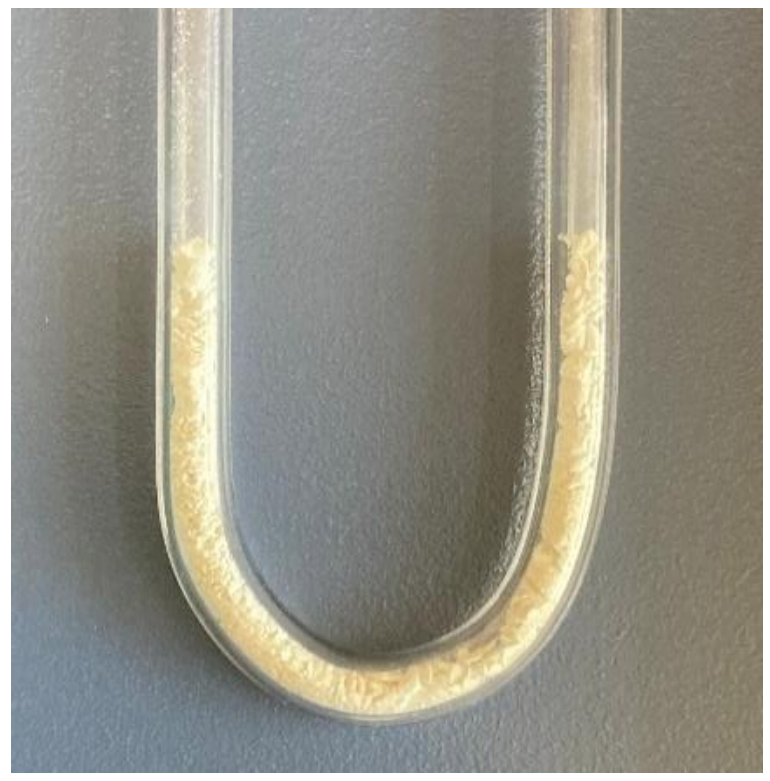

**Figure S11:** Optical image of the gel-0.25%@77K aerogel-loaded U-shaped tube.

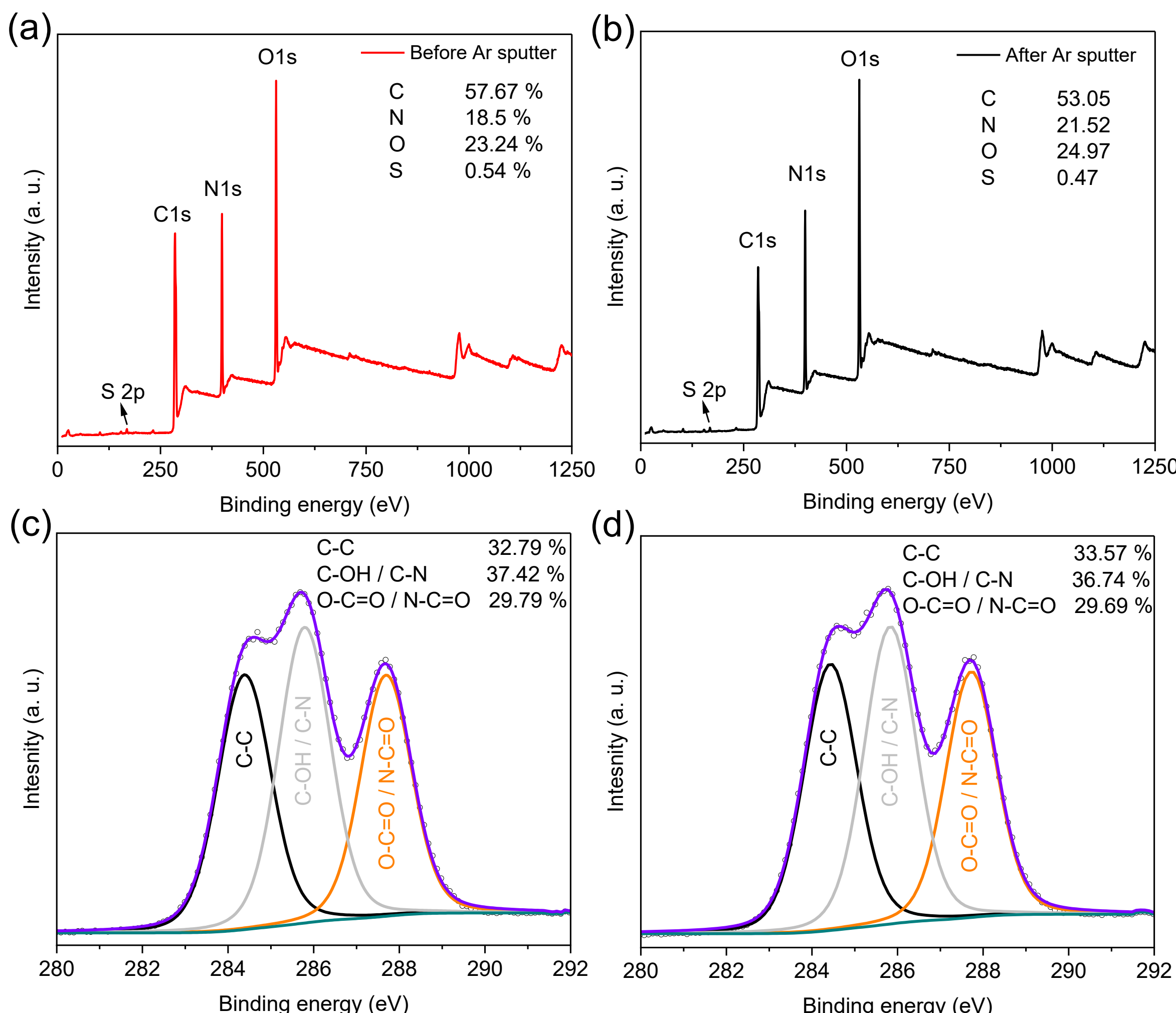

**Figure S12:** XPS survey scan of $CO_2$-adsrobed SNP (a) before and (b) after 10 s of monoatomic $Ar^+$ ion sputtering of energy 200 eV. Fitting of high resolution C1s spectra (c) before and (d) after 10 s of monoatomic $Ar^+$ ion sputtering of energy 200 eV. The symbols represent the experimental data, and the solid lines represent the fitted data.

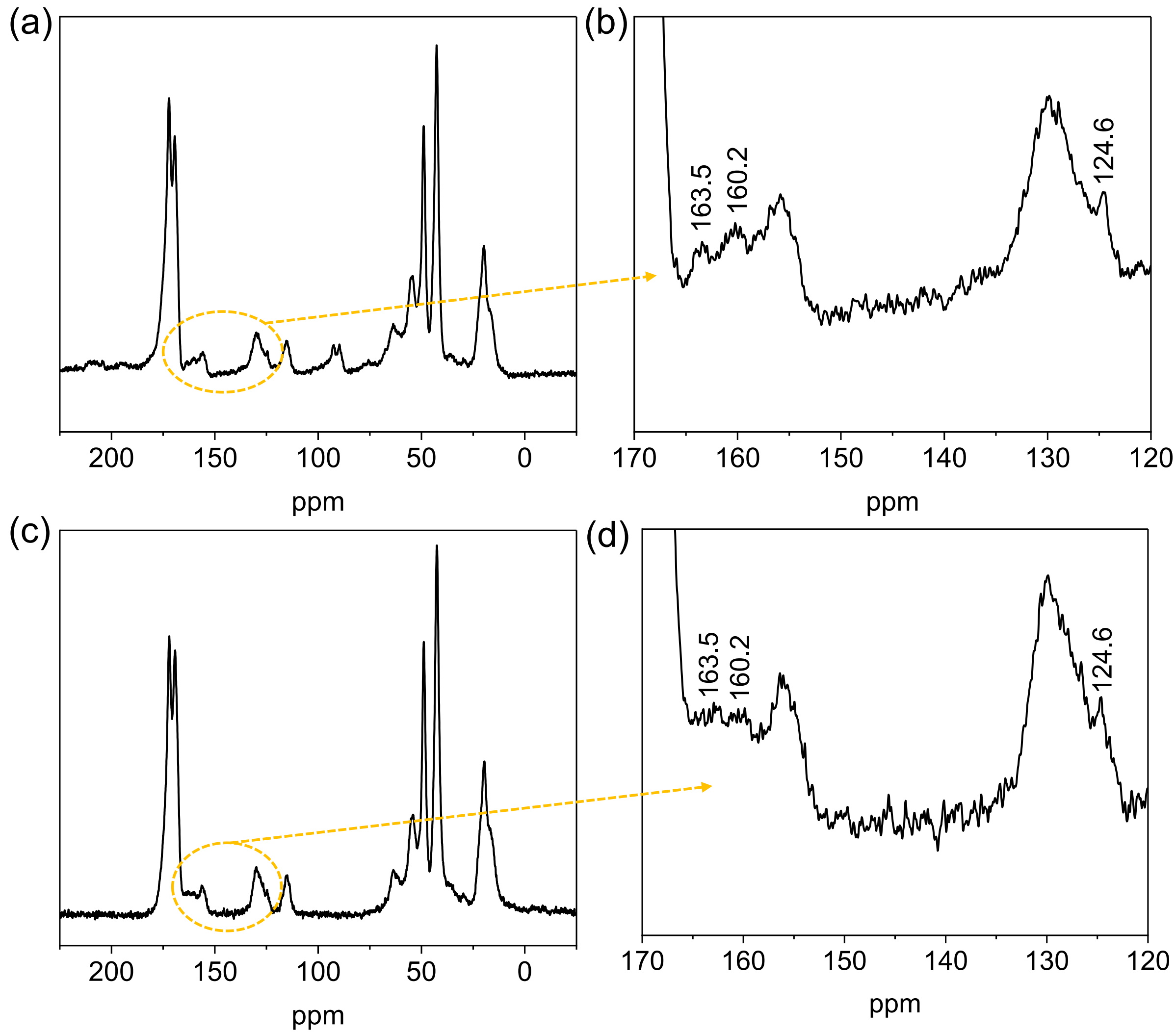

**Figure S13:** $^{13}C$ solid-state NMR spectra of silk-fibroin aerogel after $^{13}CO_2$ adsorption. Measurement was performed by cross-polarization (with continuous-wave decoupling of $^{1}H$). The magic angle spinning (MAS) rate was (a, b) 10 kHz and (c, d) 15 kHz. During NMR measurement, the sample's temperature inside the rotor was calculated to be 6.9 and 7.5±0.2 °C, respectively. The $^{13}CO_2$ gas dosing pressure was 1 bar at room temperature, though it is expected to be lower at the measured sample temperature due to pressure drop from additional adsorption and temperature drop. The number of scans was 1024 in each case.

**Table S1:** Comparison of the $CO_2$ adsorption capacity of silk-nano-fibroin-based sorbents with other high-performing amino acid and AAILs-based solid sorbents reported at nearly similar conditions.

| AAs/AAILs | Porous solid support | wt % of AA/AAILs w.r.t. sorbent | Effective surface area ($m^2$/gm) | Reported adsorption condition | Adsorption capacity (mmol/gm) | References |
|---|---|---|---|---|---|---|

| SNP | No support | Not applicable | 122 | 1 bar pure $CO_2$, 25 °C | 0.57 | This work |
|---|---|---|---|---|---|---|
| Sol-0.06%@77K | No support | Not applicable | 189 | 1 bar pure $CO_2$, 25 °C | 0.92 | This work |
| Gel-0.25%@77K | No support | Not applicable | 229 | 1 bar pure $CO_2$, 25 °C | 1.0 | This work |
| Egg white | Activated Carbon | 10 | 625 | 1 bar pure $CO_2$, 25 °C | 0.57 | Hatta et al.[14] |
| [APMIM][Lys] | Silica | 50 | 148 | 1 bar pure $CO_2$, 30 °C | 0.55 | Huang et al.[15] |
| [APMIM][Lys] | PMMA | 50 | 88 | 1 bar pure $CO_2$, 30 °C | 1.45 | Huang et al.[15] |
| [EMIM][Gly] | UiO-66 (MOF) | 5 | 1102 | 1 bar pure $CO_2$, 25 °C | 2.5 | Xia et al.[16] |
| [EMIM][Gly] | NU-1000 (MOF) | 5 | 1754 | 1 bar pure $CO_2$, 25 °C | 1.8 | Xia et al.[16] |
| Arg/PSS | PMMA | 25 | -- | 1 bar pure $CO_2$, 40 °C | 1.3 | Jiang et al.[17] |
| Sarcosine | No support | -- | -- | 1 bar pure $CO_2$, 30 °C | 2.63 | Chatterjee et al. [2] |
| Taurine | No support | -- | -- | 1 bar pure $CO_2$, 30 °C | 3.25 | Chatterjee et al.[2] |
| [apaeP$_{444}$][Lys] | Silica | 50 | 150 | 1 bar pure $CO_2$, 25 °C | 1.87 | Ren et al.[10] |
| [apaeP$_{444}$][Ala] | Silica | 50 | -- | 1 bar pure $CO_2$, 25 °C | 1.46 | Ren et al.[10] |
| [apaeP$_{444}$][Gly] | Silica | 50 | 137 | 1 bar pure $CO_2$, 25 °C | 1.46 | Ren et al. [10] |
| [apaeP$_{444}$][His] | Silica | 50 | -- | 1 bar pure $CO_2$, 25 °C | 1.46 | Ren et al.[10] |

| [apaeP444][Asp] | Silica | 50 | -- | 1 bar pure $CO_2$, 25 °C | 1.46 | Ren et al.[10] |
|---|---|---|---|---|---|---|
| [EMIM][Gly] | PMMA | 50 | -- | 1 bar pure $CO_2$, 40 °C | 1.53 | Wang et al.[12] |
| [EMIM][Ala] | PMMA | 50 | -- | 1 bar pure $CO_2$, 40 °C | 1.38 | Wang et al.[12] |
| [EMIM][Arg] | PMMA | 50 | -- | 1 bar pure $CO_2$, 40 °C | 1.01 | Wang et al.[12] |
| [EMIM][Lys] | PMMA | 50 | 27 | 1 bar pure $CO_2$, 40 °C | 1.67 | Wang et al.[12] |
| Gly | BCK-CTF | -- | 1720 | 1 bar pure $CO_2$, 25 °C | 2.61 | Dong et al.[18] |

1-aminopropyl-3-methylimidazolium lysine ([APMIM][Lys]), polystyrene sulfonate (PSS), tetramethylammonium glycinate ([N1111][Gly]), *N*-(3-aminopropyl)aminoethyl tributylphosphonium amino acid salt ([apaeP444][AA]), covalent triazine frameworks (CTFs), bis(4-cyanophenly)ketone (BCK).

**Table S2:** Comparison of differential adsorption enthalpy ($\Delta H_{ads}$) of silk-nano-fibroin-based with other state-of-the-art sorbents reported at 1 mmol $CO_2$/gm sorbent adsorption capacity.

| Sorbents | Differential adsorption enthalpy ($\Delta H_{ads}$) (-kJ/mol) | References |
|---|---|---|
| SNP | 53.1 | This work |
| Sol-0.06%@77K | 52.9 | This work |
| Gel-0.25%@77K | 53.1 | This work |
| Carbonaceous samples | 23 | Fan et al.[19] |
| Zeolite | 58 | Bae et al.[20] |
| SBA-15 (porous silica) | 20 | Mohamedali et al.[21] |
| PP1-2 (Porous polymer) | 20 | Xu et al.[22] |
| Amine@PP1-2-tren (Amine modified porous polymer) | 45 | Xu et al.[22] |
| PMMA | 44 | Huang et al.[15] |
| MNNsCya-DETA (polyamine-appended, cyanuric acid–stabilized melamine) | 53 | Mao et al.[23] |
| Amine@MOF (tetraamine-appended metal–organic frameworks) | 99 | Kim et al.[24] |
| MOF-177 (metal–organic frameworks) | 13 | Philip et al.[25] |
| [Emim][Gly]@porous silica | 87.7 | Sheshkovas et al.[7] |
| Gly@BCK-CTF (glycine-functionalized covalent triazine framework) | 33.3 | Dong et al.[3] |

| [Emim][Ala]@MOF-177 | 14 | Philip et al.[25] |
| --- | --- | --- |
| [Emim][Gly]@MOF-177 | 16 | Philip et al.[25] |
| [APMIM][Lys]@PE-SBA-15 | 25 | Huang et al.[15] |

**Table S3:** Fitting parameter of C1s core-level spectra before and after $Ar^{+}$-ion sputtering.

| Name | Before Ar sputter | | | After Ar sputter | | |
| --- | --- | --- | --- | --- | --- | --- |
| | Position (eV) | FWHM | Area % | Position (eV) | FWHM | Area % |
| C-C | 284.39 | 1.46 | 32.79 | 284.43 | 1.46 | 33.57 |
| C-OH / C-N | 285.79 | 1.43 | 37.42 | 285.83 | 1.43 | 36.74 |
| O-C=O / N-C=O | 287.69 | 1.40 | 29.79 | 287.72 | 1.40 | 29.69 |

## S1. Synthesis of silk nanoparticles from raw silk cocoon

SNP was prepared using the partial acid hydrolysis method.[26] At first, mulberry silk cocoons were cut into small pieces and boiled in a 0.05 M $Na_2CO_3$ aqueous solution for 30 min (i). After removing the outer sericin layer by boiling, the silk was washed with cold water several times and the degummed silk was dried in air. In the hydrolysis method, 0.25 gm of degummed silk was taken in a 100 ml glass beaker and 40 ml of 5 weight % $H_2SO_4$ solution was added into it. Then the beaker was placed on a hot plate and heated to 75 °C. During the acid hydrolysis process the surrounding temperature and humidity were 20-22 °C and 35-40 %, respectively. (ii). The silk dispersion was continuously stirred using a magnetic stirrer while heating. The $H_2SO_4$ weight % in the solution was increased to 15 % (7.5, 10, 12.5 and 15 %) in 4 steps by adding the equal amount of 98 weight % $H_2SO_4$ at 5 minutes interval. The solution was stirred on the hot plate for an additional 40 min. Then, the solution was placed on a sonicator bath preheated to 55 °C and sonicated for 1 hour (iii). After that, the dispersed silk was separated from the acid by centrifugation at 10000 rpm for 3 min (iv). The separated silk was taken in a glass beaker with 10 ml of water and neutralized using 0.5 M NaOH solution (v). Then the silk-fibroin was washed 3 times with DI water using centrifugation at 10000 rpm for 3 min to remove the salt generated from the acid neutralization (vi). The cleaned silk was then dispersed in water again and placed in a sonicator bath for 30 min at room temperature (vii). After sonication, the dispersion was frozen at -80 °C. Finally, the frozen dispersion was lyophilized at -48 °C to obtain the SNP (viii-ix).

## S2. Characterizations

The room temperature X-ray diffraction (XRD) patterns of the degummed silk, SNP, sol-0.06%@77K and gel-0.25%@77K were studied using a Cu-Kα X-ray diffractometer (Bruker D8 Discovery). Field emission scanning electron microscopy (FESEM) images were collected using a scanning electron microscope (SEM, Model: Zeiss 1530). Atomic force microscopy (AFM) measurement was performed using a Bruker Dimension Icon AFM. Thermo Al-Kα X-ray photoelectron spectrometer was used for the X-ray photoemission spectroscopy (XPS) study of the SNP. For the XPS and AFM studies, the SNP was dispersed in water and drop-casted on microscopic glass slides. After drop casting, the glass slides were heated in air at 90 °C for 15 min. The particle size measurement was performed using the aqueous SNP dispersion by a Zetasizer (Malvern Nano ZSP). The thermogravimetry analysis (TGA) of the SNP, sol-0.06%@77K and gel-0.25%@77K was studied from room temperature to 575 °C in $O_2$, $N_2$ and $CO_2$ gas environment using thermogravimeter (Model TGA Q550). Nitrogen adsorption desorption measurements were carried out at 77 K using a Quantachrome (Model: autosorb iQ7) instrument. The specific surface area of the silk-fibroin-based sorbents was measured by the Brunauer–Emmett–Teller (BET) method. The room temperature Fourier Transform Infrared Spectroscopy (FTIR) was studied in ambient atmosphere using an FTIR spectrometer (Thermo Fisher Scientific: Nicolet iS10). To study the FTIR spectra of SNP before $CO_2$ adsorption, it was heated in an oven at 95 °C for 30 min to desorb chemisorbed $CO_2$ occurred during air exposure, and instantly dispersed in hot water (95 °C) to avoid $CO_2$ adsorption before FTIR measurement. To study the FTIR spectra of SNP after $CO_2$ adsorption, $CO_2$-exposed sample was dispersed in water at room temperature, and its FTIR spectra was studied, and the obtained spectrum was similar to that of the air-exposed sample measured in air.

Raman spectra were collected using a Horiba LabRAM HR Evolution confocal Raman microscope. A 100 mW 532 nm laser was used as an excitation light source along with an 1800 line/mm grating spectrometer. The laser power was adjusted to 1% by a filter. A 50× long-working-distance objective with an N.A. of 0.5 was used for all measurements. All spectrum was collected under the following conditions: an acquisition time of 15 s, three accumulations, and a spectral range of 850–1200 $cm^{-1}$. The raw spectra were processed using a Python-based routine. First, baseline correction was performed using an Asymmetric Least Squares (ALS) smoothing algorithm with a smoothing parameter ($\lambda$) of $1\times10^4$, an asymmetry parameter (p) of 0.01, and 10 iterations to estimate and subtract the baseline. Subsequently, the baseline-corrected spectra were normalized and further smoothed using a Savitzky–Golay filter (window length = 11, polynomial order = 5) to reduce noise while preserving peak features.

Solid-state nuclear magnetic resonance (NMR) spectroscopy measurements were performed using the silk-fibroin aerogel before and after $CO_2$ gas adsorption. To enhance the signal-to-noise ratio and selectively track the carbon atom of the adsorbed $CO_2$ gas against the significantly larger amount of background carbon from the silk-fibroin amino acids, we used $^{13}CO_2$ gas at atmospheric pressure. To study the $^{13}C$ spectra before adsorption, the as-synthesized silk-fibroin aerogel was loaded into a Bruker 4 mm rotor, and both the sample-loaded rotor and cap (not sealed) were placed in a gas sampling bag. The gas sampling bag was partially filled with $N_2$ gas and was placed inside an oven at 90 °C for 15 minutes to release any surface-adsorbed gas or moisture. The $N_2$ gas inside the plastic bag was replaced once, and the bag was heated in the oven for 15 minutes. Then, the sample inside the bag was cooled to room temperature, and the rotor was sealed in an $N_2$ environment at atmospheric pressure using the macor cap inside the bag (with hand pressure applied from outside), without exposure to air. The sealed rotor was stored inside a gas sampling bag in the $N_2$ environment until the measurement was taken. To study the $^{13}C$ NMR spectra after $CO_2$ adsorption, we used pure $^{13}CO_2$ gas (99.0 atom% $^{13}C$, Sigma Aldrich) as the percentage of $^{13}CO_2$ in the atmosphere is very low, approximately 1%. After studying the $^{13}C$ spectra before $CO_2$ adsorption, the sample-loaded rotor and cap (separated) were placed in a gas sampling bag and heated in $N_2$ gas at 90 °C for 15 minutes. Then, the gas sampling bag was connected to a vacuum pump for 5 minutes. Finally, a small amount of $^{13}CO_2$ gas was purged into the gas sampling bag to expose the sample to $^{13}CO_2$. The rotor cap was then sealed within the bag at room temperature and atmospheric pressure, without exposure to air.

All NMR measurements were performed at 11.74 T ($^1$H frequency = 500.22 MHz and $^{13}$C frequency = 125.79 MHz) using a Bruker 4 mm probe with a magic angle spinning (MAS) rate of 10 or 15 kHz, as specified in each case. Cross-polarized magic-angle spinning (CP/MAS) $^{13}$C NMR spectra were acquired with SPINAL-64 proton decoupling (~94 kHz RF field, 7 µs $^1$H and 2.65 µs $^{13}$C 90° pulses) and a contact time of 2000 µs, averaging 1024 scans. $^1$H and $^{13}$C chemical shifts were referenced secondarily to $^1$H-adamantane at 1.85 ppm, as reported previously.[27] The measurement temperature of the sample in the rotor was set at 7±0.2 °C. For a fixed MAS rate and probe temperature, the temperature of the sample inside the rotor was calibrated using KBr-adamantane before the measurement. The sample was cooled down by cooling the bearing gas, and at each temperature, the sample was equilibrated for 15 minutes before starting the measurement. A set probe temperature of -3.1±0.2 °C at a MAS rate of 10 kHz resulted in a sample temperature of 6.9±0.2 °C. For the same sample temperature ~7.5±0.2 °C) at a 15 kHz MAS rate, the probe temperature was set to -15±0.2 °C.

## S3. $CO_2$ adsorption-desorption study

To study the $CO_2$ adsorption capacity, at first, the sorbents were heated in a vacuum at 100 °C for 1 h to remove any adsorbed $CO_2$ gas, moisture, or any other removable surface components. $CO_2$ adsorption-desorption isotherms of silk-fibroin-based sorbents were studied using a Quantachrome (Model: autosorb iQ 7) instrument. The temperature of the sample cell was controlled using a liquid bath temperature controller (Model JULABO 200F). The multi-cycle stability of the aerogels was studied by studying the $CO_2$ adsorption-desorption isotherms at 5 °C using the same Quantachrome (Model: autosorb iQ 7) instrument. To study the multi-cycle moisture stability, the aerogel gel-0.25%@77K was exposed overnight to 60 to 80% relative humidity air at around 22 °C; the next day, it was outgassed by vacuum heating at 100 °C for 30 min before the $CO_2$ adsorption-desorption isotherms measurement. The $CO_2$ adsorption capacity at 1 atm was monitored to study the stability of the aerogels.

To study the adsorption kinetics and understand the adsorption capacity in humid gas, 70 mg of the gel-0.25%@77K aerogel was loaded in a U-shaped quartz tube (**Figure S11**, ID ~ 3 mm, OD ~ 5 mm), and ~13.3 % $CO_2$ balanced $N_2$ gas was passed through the U-tube at a total flow rate of 8.3 SCCM, and the output gas from the adsorption-desorption tube was analyzed using gas chromatography (GC, Inficon Micro GC Fusion). For the study, first, the U-shaped tube was dipped in a 60 °C water bath for 10 minutes to desorb any adsorbed $CO_2$ gas. After that, the U-shaped tube was transferred to a water bath at 5 °C and kept for ~10 minutes. After the $CO_2$ adsorption step, the tube was transferred to the 60 °C water bath. During this adsorption-desorption cycle, the gas concentration in the outlet of the sample tube was monitored using the GC continuously. To study the $CO_2$ adsorption kinetics in the presence of humidity, the dry gas mixture was first passed through two conical flasks containing water to gain moisture. The relative humidity in the gas stream was measured using a humidity meter. The relative humidity in the gas stream at 5 °C was 83±2%. The humid gas was passed through the sample tube, and the adsorption kinetics were studied. To study desorption kinetics, after adsorption, the gas in the tube was switched back to dry gas, and the U-shaped tube was transferred to a hot water bath at 60 °C.

To study $CO_2$ desorption kinetics in the $CO_2$ gas environment, first, the $CO_2$ adsorption was performed by keeping the gel-0.25%@77K aerogel in a 1 atm $CO_2$ gas environment at 23 °C for 15 minutes. Then, its mass change was measured in a 1 atm $CO_2$ environment at 60 °C using TGA (Model TGA Q550).

## S4. FTIR analysis

The FTIR spectra, **Figure 3(c)**, show the absorption peaks at 3281, 3074, 2930, 1617, 1511, 1439, 1261, 1230, 1163, 1063, 976, and 693 $cm^{-1}$. The peaks at 3281 and 3074 $cm^{-1}$ are associated with the N−H symmetric and antisymmetric stretching mode vibration of the amine groups.[6,28] The peak 2930 $cm^{-1}$ is mainly associated with the aliphatic C–H stretching.[29] The broad and intense peak at 1617 $cm^{-1}$ is due to the bending vibration of N–H bond overlapped with the carbonyl bond C=O.[13] The peak at 1511 $cm^{-1}$ represents the combination of the C-N stretching mode vibration and N-H bending mode vibration in amide II.[30] The peak at 1439 is associated with the bending vibration of $CH_3$ in glycine and alanine.[4] The peak at 1165 $cm^{-1}$ is caused by the C–N stretching in tyrosine.[31] The C–N stretching vibration of glycine appears at 1063 $cm^{-1}$ which also overlaps with to the C–O bond stretching vibration mode of -OH group.[3] The peak at 693 cm-1 is associated with the at $COO^{-}$ bending.[32] Overall, the FTIR spectra show the presence of amino and carboxyl groups and suggest the presence of amino acids including glycine and alanine in the silk-fibroin-based sorbent.

# References:

(1) Babu, K. M. 11 - Silk Fibres – Structure, Properties and Applications. In *Handbook of Natural Fibres (Second Edition)*; Kozłowski, R. M., Mackiewicz-Talarczyk, M., Eds.; Woodhead Publishing Series in Textiles; Woodhead Publishing, 2020; pp 385–416. https://doi.org/10.1016/B978-0-12-818398-4.00013-X.

(2) Chatterjee, S.; Rayalu, S.; Kolev, S. D.; Krupadam, R. J. Adsorption of Carbon Dioxide on Naturally Occurring Solid Amino Acids. *Journal of Environmental Chemical Engineering* **2016**, *4* (3), 3170–3176. https://doi.org/10.1016/j.jece.2016.06.007.

(3) Dong, B.; Wang, D.-Y.; Wang, W.-J.; Tian, X.-L.; Ren, G. Post Synthesis of a Glycine-Functionalized Covalent Triazine Framework with Excellent $CO_2$ Capture Performance. *Microporous and Mesoporous Materials* **2020**, *306*, 110475. https://doi.org/10.1016/j.micromeso.2020.110475.

(4) Ouyang, H.; Guo, L.; Li, C.; Chen, X.; Jiang, B. Fabrication and Adsorption Performance for $CO_2$ Capture of Advanced Nanoporous Microspheres Enriched with Amino Acids. *Journal of Colloid and Interface Science* **2018**, *532*, 433–440. https://doi.org/10.1016/j.jcis.2018.07.121.

(5) Huang, Z.; Karami, D.; Mahinpey, N. Study on the Efficiency of Multiple Amino Groups in Ionic Liquids on Their Sorbents Performance for Low-Temperature $CO_2$ Capture. *Chemical Engineering Research and Design* **2021**, *167*, 198–206. https://doi.org/10.1016/j.cherd.2021.01.016.

(6) Wu, J.; Yang, Z.; Xie, J.; Zhu, P.; Wei, J.; Jin, R.; Yang, H. Porous Polymer Supported Amino Functionalized Ionic Liquid for Effective $CO_2$ Capture. *Langmuir* **2023**, *39* (7), 2729–2738. https://doi.org/10.1021/acs.langmuir.2c03217.

(7) Sheshkovas, A. Z.; Veselovskaya, J. V.; Rogov, V. A.; Kozlov, D. V. Thermochemical Study of $CO_2$ Capture by Mesoporous Silica Gel Loaded with the Amino Acid Ionic Liquid 1-Ethyl-3-Methylimidazolium Glycinate. *Microporous and Mesoporous Materials* **2022**, *341*, 112113. https://doi.org/10.1016/j.micromeso.2022.112113.

(8) Sun, L.; Gao, M.; Tang, S. Porous Amino Acid-Functionalized Poly(Ionic Liquid) Foamed with Supercritical $CO_2$ and Its Application in $CO_2$ Adsorption. *Chemical Engineering Journal* **2021**, *412*, 128764. https://doi.org/10.1016/j.cej.2021.128764.

(9) Uehara, Y.; Karami, D.; Mahinpey, N. Effect of Water Vapor on $CO_2$ Sorption–Desorption Behaviors of Supported Amino Acid Ionic Liquid Sorbents on Porous Microspheres. *Ind. Eng. Chem. Res.* **2017**, *56* (48), 14316–14323. https://doi.org/10.1021/acs.iecr.7b02771.

(10) Ren, J.; Wu, L.; Li, B.-G. Preparation and $CO_2$ Sorption/Desorption of N-(3-Aminopropyl)Aminoethyl Tributylphosphonium Amino Acid Salt Ionic Liquids Supported into Porous Silica Particles. *Ind. Eng. Chem. Res.* **2012**, *51* (23), 7901–7909. https://doi.org/10.1021/ie2028415.

(11) Erto, A.; Silvestre-Albero, A.; Silvestre-Albero, J.; Rodríguez-Reinoso, F.; Balsamo, M.; Lancia, A.; Montagnaro, F. Carbon-Supported Ionic Liquids as Innovative Adsorbents for $CO_2$ Separation from Synthetic Flue-Gas. *Journal of Colloid and Interface Science* **2015**, *448*, 41–50. https://doi.org/10.1016/j.jcis.2015.01.089.

(12) Wang, X.; Akhmedov, N. G.; Duan, Y.; Luebke, D.; Hopkinson, D.; Li, B. Amino Acid-Functionalized Ionic Liquid Solid Sorbents for Post-Combustion Carbon Capture. *ACS Appl. Mater. Interfaces* **2013**, *5* (17), 8670–8677. https://doi.org/10.1021/am402306s.

(13) Hiremath, V.; Jadhav, A. H.; Lee, H.; Kwon, S.; Seo, J. G. Highly Reversible $CO_2$ Capture Using Amino Acid Functionalized Ionic Liquids Immobilized on Mesoporous Silica. *Chemical Engineering Journal* **2016**, *287*, 602–617. https://doi.org/10.1016/j.cej.2015.11.075.

(14) Mohamed Hatta, N. S.; Hussin, F.; Gew, L. T.; Aroua, M. K. Enhancing Surface Functionalization of Activated Carbon Using Amino Acids from Natural Source for $CO_2$ Capture. *Separation and Purification Technology* **2023**, *313*, 123468. https://doi.org/10.1016/j.seppur.2023.123468.

(15) Huang, Z.; Mohamedali, M.; Karami, D.; Mahinpey, N. Evaluation of Supported Multi-Functionalized Amino Acid Ionic Liquid-Based Sorbents for Low Temperature $CO_2$ Capture. *Fuel* **2022**, *310*, 122284. https://doi.org/10.1016/j.fuel.2021.122284.

(16) Xia, X.; Hu, G.; Li, W.; Li, S. Understanding Reduced $CO_2$ Uptake of Ionic Liquid/Metal–Organic Framework (IL/MOF) Composites. *ACS Appl. Nano Mater.* **2019**, *2* (9), 6022–6029. https://doi.org/10.1021/acsanm.9b01538.

(17) Jiang, B.; Wang, X.; Gray, M. L.; Duan, Y.; Luebke, D.; Li, B. Development of Amino Acid and Amino Acid-Complex Based Solid Sorbents for $CO_2$ Capture. *Applied Energy* **2013**, *109*, 112–118. https://doi.org/10.1016/j.apenergy.2013.03.070.

(18) Dong, B.; Wang, D.-Y.; Wang, W.-J.; Tian, X.-L.; Ren, G. Post Synthesis of a Glycine-Functionalized Covalent Triazine Framework with Excellent $CO_2$ Capture Performance. *Microporous and Mesoporous Materials* **2020**, *306*, 110475. https://doi.org/10.1016/j.micromeso.2020.110475.

(19) Fan, X.; Zhang, L.; Zhang, G.; Shu, Z.; Shi, J. Chitosan Derived Nitrogen-Doped Microporous Carbons for High Performance $CO_2$ Capture. *Carbon* **2013**, *61*, 423–430. https://doi.org/10.1016/j.carbon.2013.05.026.

(20) Bae, T.-H.; Hudson, M. R.; Mason, J. A.; Queen, W. L.; Dutton, J. J.; Sumida, K.; Micklash, K. J.; Kaye, S. S.; Brown, C. M.; Long, J. R. Evaluation of Cation-Exchanged Zeolite Adsorbents for Post-Combustion Carbon Dioxide Capture. *Energy Environ. Sci.* **2012**, *6* (1), 128–138. https://doi.org/10.1039/C2EE23337A.

(21) Mohamedali, M.; Ibrahim, H.; Henni, A. Imidazolium Based Ionic Liquids Confined into Mesoporous Silica MCM-41 and SBA-15 for Carbon Dioxide Capture. *Microporous and Mesoporous Materials* **2020**, *294*, 109916. https://doi.org/10.1016/j.micromeso.2019.109916.

(22) Xu, C.; Bacsik, Z.; Hedin, N. Adsorption of $CO_2$ on a Micro-/Mesoporous Polyimine Modified with Tris(2-Aminoethyl)Amine. *J. Mater. Chem. A* **2015**, *3* (31), 16229–16234. https://doi.org/10.1039/C5TA01321F.

(23) Mao, H.; Tang, J.; Day, G. S.; Peng, Y.; Wang, H.; Xiao, X.; Yang, Y.; Jiang, Y.; Chen, S.; Halat, D. M.; Lund, A.; Lv, X.; Zhang, W.; Yang, C.; Lin, Z.; Zhou, H.-C.; Pines, A.; Cui, Y.; Reimer, J. A. A Scalable Solid-State Nanoporous Network with Atomic-Level Interaction Design for Carbon Dioxide Capture. *Science Advances* **2022**, *8* (31), eabo6849. https://doi.org/10.1126/sciadv.abo6849.

(24) Kim, E. J.; Siegelman, R. L.; Jiang, H. Z. H.; Forse, A. C.; Lee, J.-H.; Martell, J. D.; Milner, P. J.; Falkowski, J. M.; Neaton, J. B.; Reimer, J. A.; Weston, S. C.; Long, J. R. Cooperative Carbon Capture and Steam Regeneration with Tetraamine-Appended Metal–Organic Frameworks. *Science* **2020**, *369* (6502), 392–396. https://doi.org/10.1126/science.abb3976.

(25) Philip, F. A.; Henni, A. Incorporation of Amino Acid-Functionalized Ionic Liquids into Highly Porous MOF-177 to Improve the Post-Combustion $CO_2$ Capture Capacity. *Molecules* **2023**, *28* (20), 7185. https://doi.org/10.3390/molecules28207185.

(26) Lin, M.; Xie, W.; Cheng, X.; Yang, Y.; Sonamuthu, J.; Zhou, Y.; Yang, X.; Cai, Y. Fabrication of Silk Fibroin Film Enhanced by Acid Hydrolyzed Silk Fibroin Nanowhiskers to Improve Bacterial Inhibition and Biocompatibility Efficacy. *Journal of Biomaterials Science, Polymer Edition* **2022**, *33* (10), 1308–1323. https://doi.org/10.1080/09205063.2022.2051694.

(27) Harris, R. K.; Becker, E. D.; Menezes, S. M. C. de; Granger, P.; Hoffman, R. E.; Zilm, K. W. Further Conventions for NMR Shielding and Chemical Shifts (IUPAC Recommendations 2008). *Pure and Applied Chemistry* **2008**, *80* (1), 59–84. https://doi.org/10.1351/pac200880010059.

(28) Geminiani, L.; Campione, F. P.; Canevali, C.; Corti, C.; Giussani, B.; Gorla, G.; Luraschi, M.; Recchia, S.; Rampazzi, L. Historical Silk: A Novel Method to Evaluate Degumming with Non-Invasive Infrared Spectroscopy and Spectral Deconvolution. *Materials* **2023**, *16* (5), 1819. https://doi.org/10.3390/ma16051819.

(29) Laity, P. R.; Gilks, S. E.; Holland, C. Rheological Behaviour of Native Silk Feedstocks. *Polymer* **2015**, *67*, 28–39. https://doi.org/10.1016/j.polymer.2015.04.049.

(30) Giubertoni, G.; Caporaletti, F.; Roeters, S. J.; Chatterley, A. S.; Weidner, T.; Laity, P.; Holland, C.; Woutersen, S. In Situ Identification of Secondary Structures in Unpurified Bombyx Mori Silk Fibrils Using Polarized Two-Dimensional Infrared Spectroscopy. *Biomacromolecules* **2022**, *23* (12), 5340–5349. https://doi.org/10.1021/acs.biomac.2c01156.

(31) Koperska, M. A.; Pawcenis, D.; Bagniuk, J.; Zaitz, M. M.; Missori, M.; Łojewski, T.; Łojewska, J. Degradation Markers of Fibroin in Silk through Infrared Spectroscopy. *Polymer Degradation and Stability* **2014**, *105*, 185–196. https://doi.org/10.1016/j.polymdegradstab.2014.04.008.

(32) Ashok Kumar, R.; Ezhil Vizhi, R.; Sivakumar, N.; Vijayan, N.; Rajan Babu, D. Crystal Growth, Optical and Thermal Studies of Nonlinear Optical γ-Glycine Single Crystal Grown from Lithium Nitrate. *Optik* **2012**, *123* (5), 409–413. https://doi.org/10.1016/j.ijleo.2011.04.019.